\documentclass[aps,pre,amsmath,amssymb,lengthcheck,showpacs,superscriptaddress]{revtex4-1}
\usepackage{graphicx}
\def\vec#1{\mbox{\boldmath $#1$}}

\newcommand{\deri}[2]{{\displaystyle \frac{\partial #1 }{\partial #2 }}}
\newcommand{\derri}[2]{{\displaystyle \frac{\partial^2 #1 }{\partial #2 }}}

\begin{document}

\title{Measuring Spatial Distribution of Local Elastic Modulus in Glasses}

\author{Hideyuki Mizuno}
\email{hmizuno@ujf-grenoble.fr}
\affiliation{Laboratory for Interdisciplinary Physics, UMR 5588, Universit\'e Grenoble 1 and CNRS, 38402 Saint Martin d'H\`eres, France}

\author{Stefano Mossa}
\email{stefano.mossa@cea.fr}
\affiliation{CEA, INAC, SPrAM, UMR 5819 (UJF, CNRS, CEA), 17 rue des Martyrs, 38054 Grenoble Cedex 9, France}

\author{Jean-Louis Barrat}
\email{jean-louis.barrat@ujf-grenoble.fr}
\affiliation{Laboratory for Interdisciplinary Physics, UMR 5588, Universit\'e Grenoble 1 and CNRS, 38402 Saint Martin d'H\`eres, France}

\date{\today}

\begin{abstract}
Glasses exhibit spatially inhomogeneous elastic properties, which can be investigated by measuring their elastic moduli at a local scale.
Various methods to evaluate the local elastic modulus have been proposed in the literature.
A first possibility is to measure the \textit{local stress-local strain} curve and to obtain the local elastic modulus from the slope of the curve, 
or equivalently to use a local fluctuation formula.
Another possible route  is to assume an affine strain  and to use the applied \textit{global strain} instead of the local strain for the 
calculation of the local modulus.
Most recently a third technique has been introduced, which is easy to be implemented and has the advantage of low computational cost.
In this contribution, we compare these three approaches by using the same model glass and reveal the  differences among them caused 
by the non-affine deformations.
\end{abstract}


\maketitle

\section{Introduction}
It is well documented that glasses exhibit spatially inhomogeneous elastic properties, with coexistence of hard and soft domains when the 
elastic properties are measured at a local (typically ten atomic sizes) scale  \cite{yoshimoto_2004,makke_2011,tsamados_2009}.
The local heterogeneity in the elastic properties is reflected by the existence of a strong non-affine character in the elastic deformation of the material
\cite{wittmer_2002,tanguy_2002,leonforte_2004,leonforte_2005}: the displacement field at small scale is not obtained from the macroscopic strain, but the atoms undergo an extra relaxation described as a non-affine displacement, which has long range spatial  correlations due to the elastic character of the problem \cite{Didonna,Maloney}.
The scale $\xi$ of the elastic heterogeneities can be  assessed by measuring the local elastic properties as a function of a coarse graining size, and monitoring the convergence towards macroscopic properties \cite{tsamados_2009}.
The elastic continuum approximation for the acoustic excitations breaks down on a mesoscopic wavelength comparable to $\xi$, where a marked reduction of the sound velocity and strong scattering were observed \cite{monaco_2009,monaco2_2009}.
It has been suggested that the elastic heterogeneity is closely linked to several unusual properties of glasses, which include low-temperature thermal properties \cite{lowtem}, an excess vibrational density of states, known as the  ``Boson peak" \cite{buchenau_1984,malinovsky_1991}, and anomalous acoustic properties \cite{monaco_2009,monaco2_2009,ruffle_2003,ruffle_2006,masciovecchio_2006}. 
Theoretical models  \cite{schirmacher_2006,schirmacher_2007} have been proposed to relate the boson peak and the associated thermal and acoustic anomalies to a randomly fluctuating shear modulus.
Moreover, the localized plastic events \cite{tanguy_2006,tanguy_2010} that lead to glass plasticity are related to the elastic heterogeneities, with a strong correlation between the plastic rearrangements  and the spatial map of the local shear modulus  \cite{tsamados_2009}.
The concept of elastic heterogeneity therefore appears as a  key to understand mechanical (elastic or plastic)  properties as well as the boson peak and the associated thermal and acoustic anomalies of glasses. The quantification of these heterogeneities is of primary interest in simulation studies of amorphous systems.

Various methods have been proposed to evaluate the local elastic modulus of materials.
Probably the most natural one  is to measure the stress-strain curve at a local scale.
The local elastic modulus is then calculated from the slope (first-derivative) of the \textit{local stress} with respect to the \textit{local strain}, in the same manner as the macroscopic modulus. As for the macroscopic elastic constants, 
this approach can be  implemented either by using at a local scale the statistical mechanical formulae  \cite{lutsko_1988,cormier_2001,yoshimoto_2004} that are obtained from linear response theory, or by applying an explicit deformation to derive the local stress-local strain relation directly \cite{tsamados_2009,goldhirsch_2010}.
Another, more approximate, approach  is to assume an ``affine strain":   the applied global strain, instead of the local strain, is used  for the calculation of the local elastic modulus.
In this approach, the local modulus is calculated from the slope of the \textit{local stress-global strain} curve, so that a simple and quick stress measurement gives immediately the elastic constants.
A previous study \cite{yoshimoto_2004} reported that the local modulus calculated by this approach is qualitatively consistent with that obtained from the local stress-local strain curve, although there are some differences at a quantitative level.
Furthermore, most recently, Sollich and Barra proposed a new approach to evaluate local elastic properties \cite{sollich_2009,sollich_2012}.
In this approach,  the material is ``frozen" except for the ``target" local region.
The frozen part undergoes an  homogeneous affine deformation corresponding to an imposed external strain, and is  not allowed to relax non-affinely.
The local elastic modulus is measured from the stress response in the target local region to the applied strain.
This third approach is easy to implement and has the advantage of low computational cost. In the recent study \cite{makke_2011} this approach was used to relate cavitation in  a uniaxially strained polymer to ``weak spots" in the local bulk modulus.

The elastic modulus tensor $\vec{C}$ is composed of three components: $\vec{C} = \vec{C}^B - \vec{C}^N + \vec{C}^K$ \cite{lutsko_1988,yoshimoto_2004,lutsko_1989,wittmer2013shear}.
The first term $\vec{C}^B$ is the so-called Born term, which corresponds to the instantaneous elastic modulus under a uniform affine deformation. The second $\vec{C}^N$ is the component due to the non-affine deformation: The non-affine or internal motions of particles give rise to a decrease of the modulus. The third $\vec{C}^K$ is the contribution from the kinetic energy to the modulus, which is much smaller than the other two terms and can be neglected for dense systems.
For amorphous materials, the non-affine term $\vec{C}^N$ is an important contribution, comparable in magnitude with the affine term $\vec{C}^B$ \cite{wittmer_2002,tanguy_2002,leonforte_2005}.
The three approaches to measure the local modulus described above evaluate this non-affine component $\vec{C}^N$ in different ways.
In previous studies \cite{yoshimoto_2004,makke_2011,tsamados_2009}, all of the three different approaches appeared to provide reasonable and qualitatively consistent  values of the local moduli.
However, since these studies \cite{yoshimoto_2004,makke_2011,tsamados_2009} used different systems, the comparison remained at a qualitative level. In this contribution, we apply the three approaches on the same system and compare the corresponding results quantitatively. We point out the differences among them due to the non-affine deformations, which is discussed in details.

The paper is arranged as follows.
In Sec. \ref{moduli}, we briefly summarize the basic definitions of  local bulk and shear moduli.
In Sec. \ref{method}, our Lennard-Jones (LJ) glass system is described and the  three approaches to measure the local modulus --fully local, affine strain, and frozen matrix-- are presented. In Sec. \ref{results}, we discuss and compare the results of the local moduli obtained by using the three approaches. Finally, we conclude with a summary of our findings in Sec. \ref{conclusion}.

\section{Local bulk and shear moduli} \label{moduli}
The local elastic moduli are defined in a manner similar to the one used for macroscopic moduli. 
Let us focus on a local cubic domain $m$ of linear size $W$ in a glass sample.
The local elastic modulus in the cube $m$ can be defined as the first-derivative of the local stress with respect to the local strain.
The stress-strain relation in the domain $m$ is written  as
\begin{equation}
\vec{\sigma}^m = \vec{\sigma}^{0m} + \vec{C}^m \cdot \vec{\epsilon}^m, \label{lssr}
\end{equation}
where $\vec{\sigma}^m$, $\vec{\epsilon}^m$, and $\vec{C}^m$ are respectively the local stress, local strain, and local modulus tensors. The initial stress $\vec{\sigma}^{0m}$ generally has some non-zero values \cite{barron_1965}.
All the quantities in Eq.~(\ref{lssr}) are local ones, which depend on the position $\vec{r}$ and the size $W$ of the cube $m$.
Additionally, we remark that the local modulus tensor ${C}^m_{ijkl}$ is not necessarily symmetric with respect to the exchange of $ij$ and $kl$ like the macroscopic modulus ${C}_{ijkl}$~\footnote{By considering the existence of a strain-energy function, we obtain the symmetry of the macroscopic modulus $C_{ijkl}$, $C_{ijkl}-C_{klij}=\sigma^{0}_{kl}\delta_{ij}-\sigma^{0}_{ij}\delta_{kl}$, where $\sigma^{0}_{ij}$ is the macroscopic initial stress, and $\delta_{ij}$ is the Kronecker delta (unity when $i=j$, zero otherwise)
(see Eq. (4.20) in Ref. \cite{barron_1965}). When $\sigma^{0}_{ij}$ is isotropic, which is often true at macroscopic scale, then $C_{ijkl}=C_{klij}$.
However, such a strain-energy function does not necessarily exist at local scale.}.
Indeed, we observed that ${C}^m_{ijkl}$ is not symmetric in cubes with small $W$ for our glass system (a 3-dimensional LJ glass). Reference~\cite{tsamados_2009} also demonstrated for a 2-dimensional LJ glass that the symmetry of the local modulus tensor breaks down as the linear size $W$ becomes small.

In a 3-dimensional system there  are six mutually-independent deformations represented by the following strain tensors \cite{continuum}:
\begin{equation}
\begin{aligned}
& \vec{\epsilon}^m_b = \epsilon^m_b
\left[
\begin{array}{cccccc}
1 \\
1 \\
1 \\
0 \\
0 \\
0
\end{array}
\right],\ 
\vec{\epsilon}^m_{s1} = \epsilon^m_{s1}
\left[
\begin{array}{cccccc}
1 \\
-1 \\
0 \\
0 \\
0 \\
0
\end{array}
\right],\ 
\vec{\epsilon}^m_{s2} = \epsilon^m_{s2}
\left[
\begin{array}{cccccc}
1 \\
1 \\
-2 \\
0 \\
0 \\
0
\end{array}
\right], \\
& \vec{\epsilon}^m_{s3} = \epsilon^m_{s3}
\left[
\begin{array}{cccccc}
0 \\
0 \\
0 \\
1 \\
0 \\
0
\end{array}
\right],\ 
\vec{\epsilon}^m_{s4} = \epsilon^m_{s4}
\left[
\begin{array}{cccccc}
0 \\
0 \\
0 \\
0 \\
1 \\
0
\end{array}
\right],\ 
\vec{\epsilon}^m_{s5} = \epsilon^m_{s5}
\left[
\begin{array}{cccccc}
0 \\
0 \\
0 \\
0 \\
0 \\
1
\end{array}
\right], \label{sixdefm}
\end{aligned}
\end{equation}
where $\epsilon^m_b$, $\epsilon^m_{s1}$, $\epsilon^m_{s2}$, $\epsilon^m_{s3}$, $\epsilon^m_{s4}$, and $\epsilon^m_{s5}$ are the strengths of the strains.
The strain tensor $\vec{\epsilon}_b^m$ represents an isotropic bulk compression of the domain $m$.
The other five tensors express volume conserving shear deformations.
The two tensors, $\vec{\epsilon}_{s1}^m$ and $\vec{\epsilon}_{s2}^m$, are the ``pure" shear deformations  (plane strain and triaxial).
The three tensors, $\vec{\epsilon}_{s3}^m$, $\vec{\epsilon}_{s4}^m$, and $\vec{\epsilon}_{s5}^m$, are the ``simple" shear deformations.
All the linear deformations are realized as superpositions of these six deformations (one bulk and five shear deformations) \cite{continuum}.

From the six deformations expressed in Eq.~(\ref{sixdefm}), one  defines six moduli: one bulk modulus and five shear moduli \cite{continuum}.
The bulk modulus $K^m$ is defined from the pressure-volume change relation under the isotropic bulk deformation $\vec{\epsilon}_b^m$:
\begin{equation}
p^m = p^{0m} - K^m \frac{\delta V^m}{V^m} = p^{0m} - 3K^m \epsilon^m_b. \label{mmbulk}
\end{equation}
Here the pressure $p^m$ is the trace of the stress tensor, $p^m = -(\sigma_{xx}^m+\sigma_{yy}^m+\sigma_{zz}^m)/3$, and the volume change $\delta V^m$ is written as $\delta V^m/V^m = \epsilon_{xx}^m+\epsilon_{yy}^m+\epsilon_{zz}^m =3 \epsilon_b^m$.
In addition, five shear moduli, $G^m_1$, $G^m_2$, $G^m_3$, $G^m_4$, and $G^m_5$, are defined from the shear stress-shear strain relations under two pure shear deformations, $\vec{\epsilon}_{s1}^m$ and $\vec{\epsilon}_{s2}^m$, and three simple shear deformations, $\vec{\epsilon}_{s3}^m$, $\vec{\epsilon}_{s4}^m$, and $\vec{\epsilon}_{s5}^m$, respectively:
\begin{equation}
\begin{aligned}
\sigma^m_{s1} &= \sigma_{s1}^{0m} + 2 G^m_1 \epsilon^m_{s1}, \qquad
\sigma^m_{s2}  = \sigma_{s2}^{0m} + 2 G^m_2 \epsilon^m_{s2}, \\
\sigma^m_{s3} &= \sigma_{s3}^{0m} + 2 G^m_3 \epsilon^m_{s3}, \qquad
\sigma^m_{s4}  = \sigma_{s4}^{0m} + 2 G^m_4 \epsilon^m_{s4}, \\
\sigma^m_{s5} &= \sigma_{s5}^{0m} + 2 G^m_5 \epsilon^m_{s5}.
\end{aligned} \label{ssrelation}
\end{equation}
Here, the five shear stresses are written as $\sigma_{s1}^m = (\sigma_{xx}^m-\sigma_{yy}^m)/2$, $\sigma_{s2}^m = (\sigma_{xx}^m+\sigma_{yy}^m-2\sigma_{zz}^m)/4$, $\sigma_{s3}^m = \sigma_{xy}^m$, $\sigma_{s4}^m = \sigma_{xz}^m$, and $\sigma_{s5}^m = \sigma_{yz}^m$, respectively.
These bulk and shear moduli can be expressed as linear combinations of $C_{ijkl}^m$:
\begin{equation}
\begin{aligned}
3K^m &= (C^m_{xxxx}+C^m_{yyyy}+C^m_{zzzz}+C^m_{xxyy}+C^m_{yyxx} \\
  & \quad \ +C^m_{xxzz}+C^m_{zzxx}+C^m_{yyzz}+C^m_{zzyy})/3, \\
2G^m_1 & = (C^m_{xxxx}+C^m_{yyyy}-C^m_{xxyy}-C^m_{yyxx})/2, \\
2G^m_2 & = (C^m_{xxxx}+C^m_{yyyy}+4C^m_{zzzz}+C^m_{xxyy}+C^m_{yyxx} \\
    &  \quad -2C^m_{xxzz}-2C^m_{zzxx}-2C^m_{yyzz}-2C^m_{zzyy})/6, \\
2G^m_3 & = C^m_{xyxy}, \\
2G^m_4 & = C^m_{xzxz}, \\
2G^m_5 & = C^m_{yzyz},
\end{aligned} \label{transform}
\end{equation}
which correspond to the diagonal components of the matrix $\hat{\vec{C}}^m = {\vec{P}^{m}}^{-1} \vec{C}^m \vec{P}^m$ with $\vec{P}^m = [\vec{\epsilon}^m_b,\vec{\epsilon}^m_{s1},\vec{\epsilon}^m_{s2},\vec{\epsilon}^m_{s3},\vec{\epsilon}^m_{s4},\vec{\epsilon}^m_{s5}]$.
In the case of isotropic systems, like a LJ glass, the macroscopic modulus tensor $\vec{C}$ is characterized by one bulk modulus $K = \lambda + 2\mu/3$ and one shear modulus $G=\mu$ \cite{continuum,barrat_1988}, where $\lambda$ and $\mu$ are the Lam$\acute{\text{e}}$ constants.
As the size $W$ of the cube $m$ becomes large, the local bulk modulus $K^m$ tends to the macroscopic value of $K$, and the five shear moduli, $G_1^m$, $G_2^m$, $G_3^m$, $G_4^m$, and $G_5^m$, which generally have different values from each other, converge to the same value of the macroscopic $G$.

\section{Methods} \label{method}

\subsection{System preparation}
The system considered in the present study is a 3-dimensional LJ monatomic glass model, described in Ref. \cite{monaco2_2009}.
The interaction energy between two particles is $\phi(r)= 4 \epsilon [ (\sigma/r)^{12}-(\sigma/r)^{6}]$, where $r$ is the distance between the two particles,  $\epsilon$ is the depth of the potential well, and $\sigma$ is the particle diameter.
The potential was cut-off and shifted to zero at $r = r_c = 2.5 \sigma$~\footnote{The Born term $C_{ijkl}^{Bm}$ included in the calculation of the modulus tensor in Eq. (\ref{ffmodulus}), involves the second derivative of the potential. Due to the truncation of the potential, this gives an impulsive correction to the Born term~\cite{Xu2012impulsive}. We have checked that this correction is always negligible in our system.}.
The system contains $N=4,000$ particles in a simulation box of constant volume $V$ under periodic boundary conditions.
In the following, all numerical values are expressed in LJ units: length in $\sigma$, temperature in $\epsilon/k_B$ ($k_B$ is the Boltzmann constant), and time in $\tau=(m \sigma^2/\epsilon)^{1/2}$ ($m$ is the mass of a particle).
The number density was fixed at $\hat{\rho} = N/V=1.015$, which implies a linear dimension of the simulation box $L=V^{1/3}=15.8$.
For reference, we note that at the  number density considered here, $\hat{\rho} =1.015$, the melting temperature of the system is $T_m \simeq 1$, and the glass transition temperature is $T_g \simeq 0.4$ \cite{robles_2003}.
The glass phase of the system was realized as follows.
We first equilibrated the system at the temperature $T=2$ in the normal liquid phase by using a standard $NVT$ molecular dynamics (MD) simulation. Next, we quenched the system down to $T=10^{-3}$ in the glass phase with a fast quench rate $dT/dt = 4 \times 10^2$. After quenching, the system was relaxed for a sufficiently long time to stabilize the total energy.

\subsection{Three approaches to measure local elastic modulus} \label{threemethod}
We subdivided the cubic glass sample (box size $L=15.8$) into a grid of size $M \times M \times M$ and measured the local elastic modulus for small cubic domains (size $W$) centered at these grid points.
In the present study, we set the cube size to be $W=3.16(=L/5),\ 5.27(=L/3)$, and $7.90(=L/2)$, and the grid size is set to be $40 \times 40 \times 40$ for $W=3.16$ and $20 \times 20 \times 20$ for $W=5.27$ and $7.90$.
The small cubes are distinguished by the index $m$ $(m=1,2,...,M^3)$.
In the following, we describe the three approaches used to measure the local modulus.

\begin{figure*}[t]
\includegraphics[scale=1, angle=0]{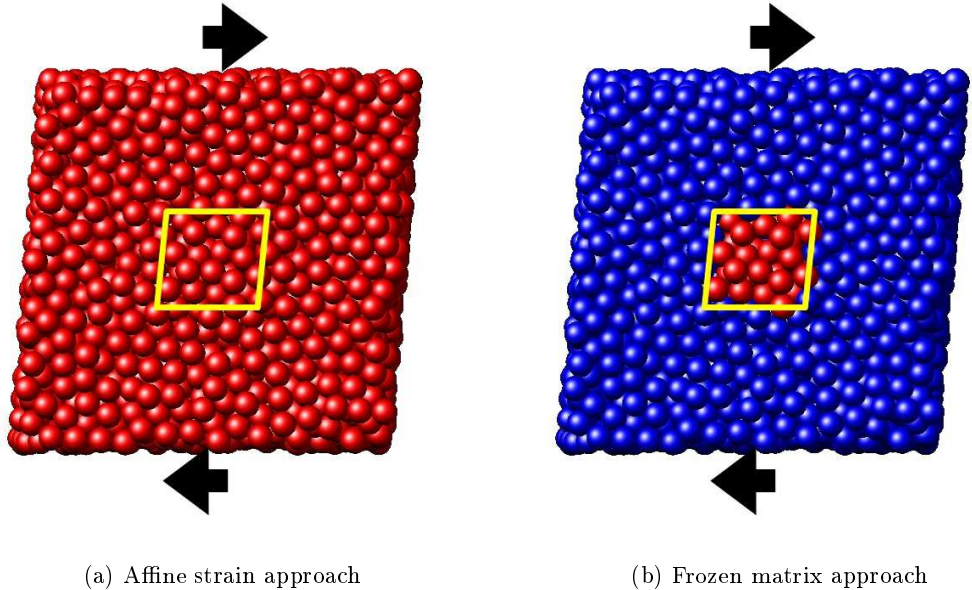}
\vspace*{0mm}
\caption{(Color online)
Schematic illustration of the simple shear deformation: (a) affine strain approach and (b) frozen matrix approach.
The cubic box drawn by the thick (yellow) lines indicates the local cube $m$. In the affine strain approach (a), all the particles (red particles) are allowed to move non-affinely. In the frozen matrix approach (b), only the particles in the local cube $m$ (red particles) can move freely, whereas the particles in the other frozen part (blue particles) are restricted to only move affinely.
} \label{illust_fig}
\end{figure*}

\subsubsection{Fully local approach} \label{threemethod1}
In this approach, one considers  the local stress $\vec{\sigma}^m$ and the local strain $\vec{\epsilon}^m$. The local modulus $\vec{C}^m$ is obtained from the first-derivative of $\vec{\sigma}^m$ with respect to $\vec{\epsilon}^m$, as in Eq. (\ref{lssr}). The approach can be implemented either by using the equilibrium fluctuation formula \cite{lutsko_1988,cormier_2001,yoshimoto_2004} or by performing an explicit deformation to obtain the local stress-local strain relation directly \cite{tsamados_2009,goldhirsch_2010}.
In the present study, the fluctuation formula was used, as described below.

Fluctuation formulae are obtained within the framework of equilibrium  statistical mechanics \cite{lutsko_1988,cormier_2001,yoshimoto_2004,lutsko_1989,squirey_1968,ray_1984,ray_1985,wittmer2013shear}.
The local stress $\sigma_{ij}^m$ for the small cube $m$ is calculated as
\begin{equation}
\sigma_{ij}^m = -\hat{\rho}^m T \delta_{ij} + \frac{1}{W^3} \sum_{a<b} \deri{\phi(r^{ab})}{r^{ab}} \frac{r^{ab}_i r^{ab}_j}{r^{ab}}\frac{q^{ab}}{r^{ab}}, \label{ffstress}
\end{equation}
where $\hat{\rho}^m$ is the number density in the cube $m$, $r^{ab}_i$ denotes the vector joining particles $a$ and $b$, and $r^{ab}$ is the distance between these two particles.
The quantity $q^{ab}$ represents the length of the line segment $r^{ab}_i$ located inside the cube $m$.
If the vector $r^{ab}_i$ is located outside the cube $m$, then $q^{ab}=0$.
The term $q^{ab}/r^{ab}$ determines the contribution of each pairwise interaction to the local stress $\sigma_{ij}^m$.
The summation of $\sigma_{ij}^m$ over the entire system yields the usual macroscopic stress tensor $\sigma_{ij}$ \cite{simpleliquid}:
\begin{equation}
\begin{aligned}
\sigma_{ij} & = \frac{1}{V} \sum_{m} W^3 \sigma^m_{ij} \\
            & = -\hat{\rho} T \delta_{ij} + \frac{1}{V} \sum_{a<b} \deri{\phi(r^{ab})}{r^{ab}} \frac{r^{ab}_i r^{ab}_j}{r^{ab}}. \label{macrostress}
\end{aligned}
\end{equation}

The local modulus tensor $C_{ijkl}^m$ for the cube $m$ is calculated from the following equations:
\begin{equation}
\begin{aligned}
C_{ijkl}^m & = C_{ijkl}^{Bm} - C_{ijkl}^{Nm} + C_{ijkl}^{Km}, \\
C_{ijkl}^{Bm} & = \frac{1}{W^3} \left< \sum_{a<b} \left( \derri{\phi}{{r^{ab}}^2} - \frac{1}{r^{ab}}\deri{\phi}{r^{ab}} \right) \frac{r^{ab}_i r^{ab}_j r^{ab}_k r^{ab}_l}{ {r^{ab}}^2 }\frac{q^{ab}}{r^{ab}} \right>, \\
C_{ijkl}^{Nm} & = \frac{V}{T} [\left< \sigma^m_{ij} \sigma_{kl} \right>-\left< \sigma^m_{ij} \right>\left< \sigma_{kl} \right>], \\
C_{ijkl}^{Km} & = 2\left< \hat{\rho}^m \right>T (\delta_{ik} \delta_{jl} + \delta_{il}\delta_{jk}),
\end{aligned} \label{ffmodulus}
\end{equation}
where $\left< \right>$ represents the ensemble average~\footnote{The Eq. (\ref{ffmodulus}) (fluctuation formula) is formulated from the second derivative of the local energy density with respect to the local Green-Lagrange strain tensor \cite{lutsko_1988,yoshimoto_2004}.
When the initial stress $\vec{\sigma}^{0m}$ has finite values, a deviation from the definition of Eq. (\ref{lssr}) (the first derivative of the stress with respect to the strain) appears.
This deviation is taken care of by adding the correction term $C_{ijkl}^{Cm}$ to the modulus $C_{ijkl}^m$ in Eq.~(\ref{ffmodulus}) (see Eq.~(4.19) in Ref.~\cite{barron_1965}):
\begin{equation}
\begin{aligned}
C_{ijkl}^{Cm} & = -\frac{1}{2} \big( 2\left<\sigma^m_{ij}\right>\delta_{kl}-\left<\sigma^m_{ik}\right>\delta_{jl}-\left<\sigma^m_{il}\right>\delta_{jk} \\
              & \qquad  \quad -\left<\sigma^m_{jk}\right>\delta_{il}-\left<\sigma^m_{jl}\right>\delta_{ik} \big).
\end{aligned} \label{ffcorrect}
\end{equation}
In this study we included the correction term $C_{ijkl}^{Cm}$ into the Born term $C_{ijkl}^{Bm}$.
We have checked that the present system has small values of the initial stress $\vec{\sigma}^{0m}$, leading to a small contribution $C_{ijkl}^{Cm}$.}.
As described in the introduction, there are three contributions to the elastic constants: the affine component (the Born term) $C_{ijkl}^{Bm}$, the non-affine component $C_{ijkl}^{Nm}$~\footnote{By substituting $\sigma_{ij}=(1/V)\sum_m W^3 \sigma^m_{ij}$ in $C_{ijkl}^{Nm}$ of Eq.~(\ref{ffmodulus}), the non-affine component $C_{ijkl}^{Nm}$ is written as
\begin{equation}
C_{ijkl}^{Nm} = \sum_{n} \frac{W^3}{T} [\left< \sigma^m_{ij} \sigma^n_{kl} \right>-\left< \sigma^m_{ij} \right>\left< \sigma^n_{kl} \right>],
\end{equation}
from which we can consider $C_{ijkl}^{Nm}$ as the sum of correlations of local stress fluctuations.}, and the kinetic contribution $C_{ijkl}^{Km}$.
We note that the non-affine term $C_{ijkl}^{Nm}$, calculated through the correlation between $\sigma_{ij}^m$ and $\sigma_{kl}$, is not symmetric with respect to the exchange of $ij$ and $kl$, and so the total modulus $C_{ijkl}^m$ is not a symmetric tensor in general.
Earlier work \cite{yoshimoto_2004} measured the local shear modulus of a polymer glass by using this method and revealed the inhomogeneous distribution of the modulus.

In order to apply Eqs. (\ref{ffstress})-(\ref{ffmodulus}),  the averages were performed on  trajectories generated by   standard $NVT$ MD simulation on the glass sample at the low temperature $T=10^{-3}$.
The time step of  the MD simulation was  $\delta t = 5 \times 10^{-3}$, and the total duration of the runs  was $t=10^5$ ($2 \times 10^7$ steps). The averages were performed over  $10^4$ configurations, separated by a time lag of $10$ LJ units.

\begin{figure*}[t]
\includegraphics[scale=1, angle=0]{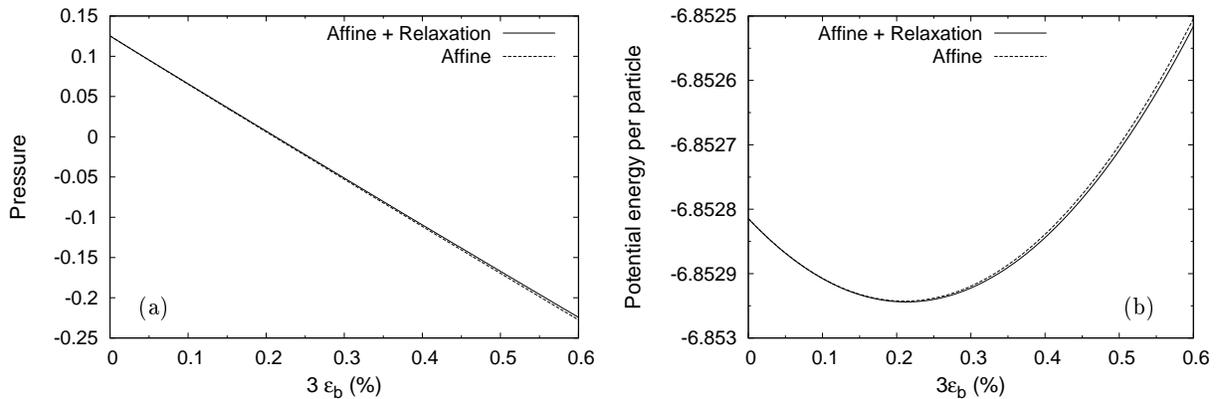}
\vspace*{0mm}
\caption{(a) Pressure and (b) potential energy per particle versus strain $3\epsilon_b$, under the quasi-static bulk deformation $\vec{\epsilon}_{b}$. We show two deformation curves: affine deformation with relaxation (solid curve) and affine deformation without relaxation (dashed curve). The macroscopic bulk modulus obtained from the slope of the stress-strain curve is $K=59.7$ (Affine + Relaxation) and $K^B=60.2$ (Affine). We also obtained the same values of $K$ and $K^B$ from a quadratic fit of the potential energy.
}
\label{mbulk}
\end{figure*}

\subsubsection{Affine strain approach} \label{threemethod2}
In this second approach,  one assumes an  ``affine strain", i.e., that the entire glass sample is strained uniformly.
This assumption says that the local strains $\vec{\epsilon}^m$ of all the small cubes $m$ are represented by the global strain $\vec{\epsilon}$ applied to the sample.
The local modulus $\vec{C}^m$ is defined and calculated based on Eqs. (\ref{lssr})-(\ref{transform}), with the local $\vec{\epsilon}^m$ replaced by the global $\vec{\epsilon}$.
This approach, which is obviously very simple to implement, ignores the spatial variations of the local strain field or, equivalently, its non-affine component.

To implement this approach we deform the entire glass sample in six ways: one bulk deformation and five shear deformations (two pure shear and three simple shear deformations), which correspond to the six strain tensors in Eq. (\ref{sixdefm}) with $\vec{\epsilon}^m$ replaced by $\vec{\epsilon}$.
The simple shear deformation is illustrated schematically in Fig. \ref{illust_fig} (a).
For each deformation, we calculate the local stress $\sigma_{ij}^m$ of the cube $m$ as a function of the applied global strain $\epsilon$. Here, we use a formulation for ${\sigma}^m_{ij}$ slightly different from that of Eq. (\ref{ffstress}) used in the fully local approach: ${\sigma}^m_{ij}$ is calculated from the summation of the atomic stresses over the cubic domain $m$ \cite{irving_1950,macneill_2010},
\begin{equation}
\sigma_{ij}^m = -\hat{\rho}^m T \delta_{ij} + \frac{1}{2 W^3} \sum_{a \in m} \sum_{b=1, b \neq a}^N \deri{\phi(r^{ab})}{r^{ab}} \frac{r^{ab}_i r^{ab}_j}{r^{ab}}, \label{gsstress}
\end{equation}
where the summation of $a$ is performed over particles in the cube $m$~\footnote{The exact statistical mechanics derivation allows one to establish Eq.(\ref{ffstress}) for the local stress tensor, which exactly conserves momentum. Note that  here the quantity $q^{ab}$ is exactly the length of the line segment $r_i^{ab}$ located inside the cube. In particular, it also accounts for the cases where both $a$ and $b$ are outside the cube. In contrast, Eq.(\ref{gsstress}) equally shares the contribution to the summation between the two atoms. We have chosen to consider this simplified formulation for computational purposes and checked that corrections to the exact form are negligible.}.
The same formulation of ${\sigma}^m_{ij}$ was used in Refs. \cite{makke_2011,macneill_2010}.
We note that the definition Eq. (\ref{gsstress}) of ${\sigma}^m_{ij}$ corresponds to the Eq. (\ref{ffstress}) with the term $q^{ab}/r^{ab}$,
\begin{equation}
\begin{aligned}
\frac{q^{ab}}{r^{ab}} = 
\begin{cases}
\ 1 & (a,b \in m), \\
\ 1/2 & (a \in m, b \notin m).
\end{cases}
\end{aligned}
\end{equation}
The usual macroscopic stress tensor $\sigma_{ij}$ is still recovered by the summation of $\sigma_{ij}^m$ over the entire system, as in Eq. (\ref{macrostress}). The bulk modulus $K^m$ is now obtained from the local pressure-global volume change relation, i.e., Eq. (\ref{mmbulk}) with the local $\epsilon^m_b$ replaced by the global $\epsilon_b$.
The shear moduli, $G_1^m$, $G_2^m$, $G_3^m$, $G_4^m$, and $G_5^m$, are calculated from the local shear stress-global shear strain relations, i.e., Eq. (\ref{ssrelation}) with the local $\epsilon^m_s$ replaced by the global $\epsilon_s$.
According to the study \cite{yoshimoto_2004}, the assumption of ``affine strain" can be qualitatively acceptable but causes quantitative deviations from the fully local  approach. We examine in details the validity of this second approach in the following.

In the present study, we have deformed the glass sample by using a ``quasi-static" protocol, i.e., the system was deformed at zero temperature $T=0$, as described in Refs. \cite{tanguy_2002,tanguy_2006,tsamados_2009}.
To achieve this, the system was first quenched using a steepest descent method from $T=10^{-3}$ to $T=0$, into the nearest energy minimum.
Next, the system was submitted to an imposed deformation by applying strain steps $\delta \epsilon = 10^{-5}$ with Lees-Edwards periodic boundary conditions \cite{Allen1986}.
After each strain step $\delta \epsilon$ was imposed, the entire system was relaxed into its new closest energy minimum by the steepest descent method. The quasi-static deformation was performed until the applied strain reached $\epsilon=0.002$ (i.e., $0.2 \%$).
We note that $\epsilon=0.002$ is small enough that the system deforms elastically, and the stress-strain curve is linear for all of the studied six deformations.
We also remark that the difference of the temperature between $T=0$ of this approach and $T=10^{-3}$ of the fully local approach is very small so that this temperature difference is not expected  to cause a discrepancy  between those two calculations.

\begin{figure*}[t]
\includegraphics[scale=1, angle=0]{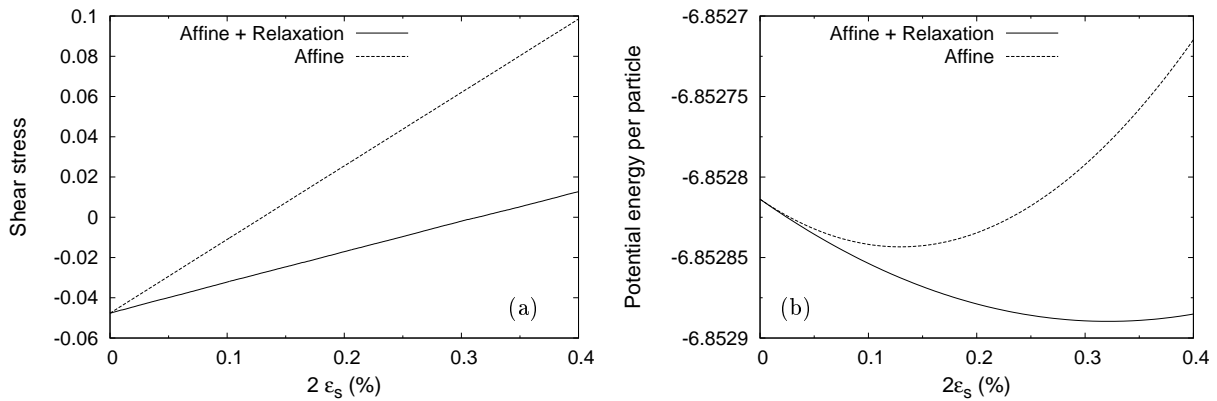}
\vspace*{0 mm}
\caption{(a) Shear stress and (b) potential energy per particle versus strain $2\epsilon_{s}$ under the quasi-static shear deformation $\vec{\epsilon}_{s3}$.
We show two deformation curves: affine deformation with relaxation (solid curve) and affine deformation without relaxation (dashed curve).
The macroscopic shear modulus obtained from the slope of the stress-strain curve is $G=14.9$ (Affine + Relaxation) and $G^B=36.5$ (Affine).
We also obtained the same values of $G$ and $G^B$ from the quadratic fitting of the potential energy.
Note that we obtained five values of $G \simeq 15$ and $G^B \simeq 36$ from the five shear deformations $\vec{\epsilon}_{s1}$, $\vec{\epsilon}_{s2}$, $\vec{\epsilon}_{s3}$, $\vec{\epsilon}_{s4}$, and $\vec{\epsilon}_{s5}$.
The values of $G=14.9$ and $G^B=36.5$ are the averages over these values.
}
\label{mshear}
\end{figure*}

\subsubsection{Frozen matrix approach} \label{threemethod3}
The third approach was originally introduced by Sollich and Barra \cite{sollich_2009,sollich_2012}.
This also is an explicit type method, similar to the  affine strain approach~\footnote{The frozen matrix approach can be also implemented by using the fluctuation formula inside a cube, with frozen boundaries. We have confirmed that the fluctuation formula and the explicit deformation produce identical results for the shear modulus. For the bulk modulus, large pressure fluctuations result in a lower value when using the fluctuation formula with frozen boundaries.}.
We first ``freeze" the system except for the ``target" local region, i.e., the local cube $m$.
The frozen region is restricted to deform only affinely due to the external strains, not allowed to relax non-affinely, whereas the target region $m$ deforms non-affinely.
Then we apply six types of deformations, one bulk and five shear deformations, on the entire sample.
We show a schematic illustration of this approach in Fig.~\ref{illust_fig}(b), where the red particles are particles in the cube $m$, and the blue particles are particles in the frozen region.
During the deformations, the particles in the cube $m$ can move freely, whereas the particles in the frozen region are displaced only affinely corresponding to the external strains.
In this situation, where the parts surrounding the local cube $m$ are frozen, the local strain $\vec{\epsilon}^m$ of the cube $m$ coincides exactly with the applied global strain $\vec{\epsilon}$.
For each deformation, we calculate the local stress ${\sigma}^m_{ij}$ of the cube $m$ as the function of the local strain ${\epsilon}^m={\epsilon}$.
We note that the formulation of Eq. (\ref{gsstress}) (i.e., the summation of the atomic stresses) is used for the calculation of $\sigma_{ij}^m$.
Then, one bulk modulus $K^m$ and five shear moduli, $G_1^m$, $G_2^m$, $G_3^m$, $G_4^m$, and $G_5^m$, are determined from Eqs. (\ref{mmbulk}) and (\ref{ssrelation}).
This approach is also easily implemented, as   the local strain $\vec{\epsilon}^m = \vec{\epsilon}$ is an input quantity.
A recent numerical work \cite{makke_2011} investigated the local bulk modulus of a glassy polymer by using this approach and obtained reasonable results in relation to cavitation events.

We used a ``quasi-static" protocol, as  in the affine strain approach.
We first quenched the glass sample from $T=10^{-3}$ to $T=0$ using a steepest descent method, and then the quenched system was frozen except for the local cube $m$.
The frozen system was submitted to an imposed deformation by applying strain steps $\delta \epsilon = 10^{-5}$, where all the particles move affinely.
After each strain step $\delta \epsilon$ was imposed, the system was relaxed by the steepest descent method.
During the relaxation (energy minimization), only the particles in the target cube $m$ (red particles in Fig. \ref{illust_fig} (b)) can move toward the minimum energy point, whereas the particles in the frozen part (blue particles in Fig. \ref{illust_fig} (b)) are stuck.
The deformation was performed until the applied strain reaches $\epsilon=\epsilon^m=0.002$ (i.e., $0.2 \%$).

\section{Results} \label{results}

\subsection{Macroscopic elastic modulus}
We first investigated the macroscopic stress-strain relation and the macroscopic modulus, which are obtained from the quasi-static deformation. In Figs. \ref{mbulk} and \ref{mshear}, we plot the macroscopic stress as the function of the applied global strain: Fig. \ref{mbulk} for the pressure $p$ under the isotropic bulk deformation $\vec{\epsilon}_b$, and Fig. \ref{mshear} for the shear stress $\sigma_{s3}$ under the simple shear deformation $\vec{\epsilon}_{s3}$.
The values of the bulk modulus $K$ and the shear modulus $G$, which were obtained from the slopes of the curves, are $K=59.7$ and $G=14.9$.
Note that we obtained five values of $G \simeq 15$ from the five shear deformations $\vec{\epsilon}_{s1}$, $\vec{\epsilon}_{s2}$, $\vec{\epsilon}_{s3}$, $\vec{\epsilon}_{s4}$, and $\vec{\epsilon}_{s5}$.
The value $G=14.9$ is the average over these five values.
The values of $K=59.7$ and $G=14.9$ are consistent with the previous work \cite{monaco2_2009}, where the longitudinal and transverse sound speeds, $c_L= \sqrt{(K+4G/3)/\rho} \simeq 8.8$ and $c_T= \sqrt{G/\rho} \simeq 3.8$ ($\rho=1.015$ is the mass density), were obtained for the same glass model as ours.
In addition, in Figs. \ref{mbulk} and \ref{mshear} we also plot the potential energy per particle.
The potential energy $\Phi$ is changed  by the stress, according to $d \Phi = -P Vd(3\epsilon_b) $ during compression  and $d \Phi=\sigma_{s} Vd(2\epsilon_{s})$ during shear.
Since the stress is a linear function of the strain, the potential energy is a quadratic function of the strain as shown in Figs. 2(b) and 3(b).
We obtained the same values of $K \simeq 60$ and $G \simeq 15$ from the quadratic fitting of the potential energy.

Also, Figs. \ref{mbulk} and \ref{mshear} display  the stress and the potential energy for the affine deformation, where the relaxation (energy minimization) is not performed.
For the isotropic bulk deformation in Fig. \ref{mbulk}, the pressure profile is almost the same as that with relaxation.
The bulk modulus $K^B=60.2$ (Born term) of the affine deformation is nearly equal to $K=59.7$.
The potential energy also shows the same profile in both cases, with and without relaxation, respectively.
This result indicates that the non-affine component is very small, and the bulk modulus is dominantly determined by the affine component.
A  previous study \cite{leonforte_2005} obtained  similar results for a slightly polydisperse LJ glass.
On the other hand, it is clearly observed that the relaxation causes a marked decrease of the shear stress and the potential energy under the shear deformation.
The shear modulus $G^B=36.5$ (Born term) of the affine deformation is much higher than $G=14.9$.
The non-affine component $G^N=G^B-G=21.6$ is of the same order magnitude as the affine component \cite{leonforte_2005,tanguy_2002,wittmer_2002}.
Therefore, the shear modulus has a large non-affine component, which is important for amorphous materials.

\begin{figure*}[p!]
\includegraphics[scale=1, angle=0]{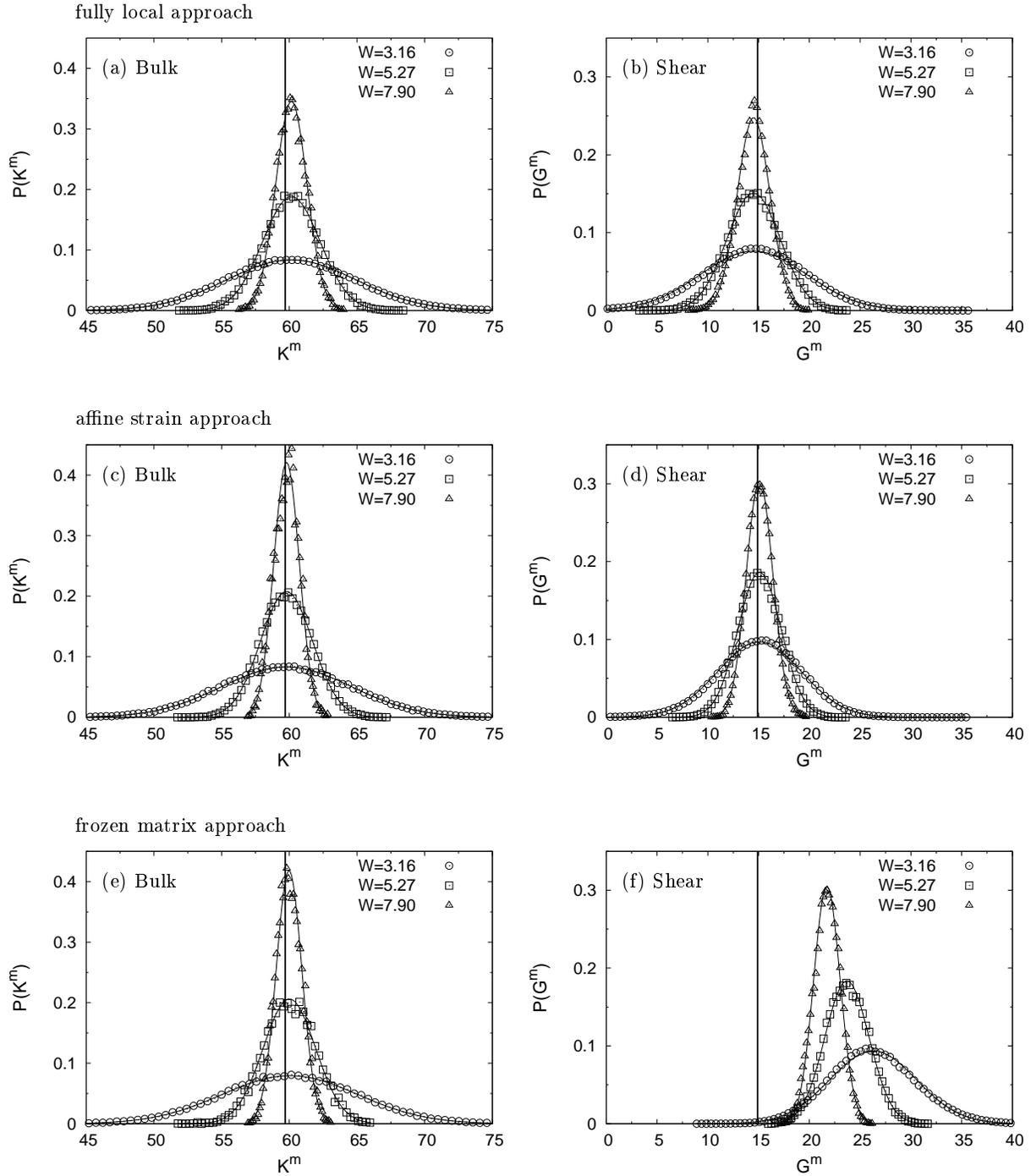}
\vspace*{0mm}
\caption{Distributions of bulk modulus and shear modulus for different cube sizes $W=3.16$, $5.27$, and $7.90$ calculated by the three approaches described in the text: fully local approach ((a),(b)), affine strain approach ((c),(d)), and frozen matrix approach ((e),(f)).
The shear modulus distribution $P(G^m)$ is obtained by averaging the five distributions $P(G_1^m)$, $P(G_2^m)$, $P(G_3^m)$, $P(G_4^m)$, and $P(G_5^m)$.
We also show the Gaussian distribution functions fitted to each distribution (solid lines).
The vertical solid lines indicate the values of the macroscopic moduli obtained from Figs. \ref{mbulk} and \ref{mshear}, i.e., $K = 59.7$ and $G = 14.9$.
In (a)-(e), the average value is independent of the cube size $W$ and coincides with the macroscopic value, whereas in (f), the average value varies with $W$ and seems to tend to the macroscopic value with increasing $W$.} \label{m_dis}
\end{figure*}

\begin{figure*}[t]
\includegraphics[scale=1, angle=0]{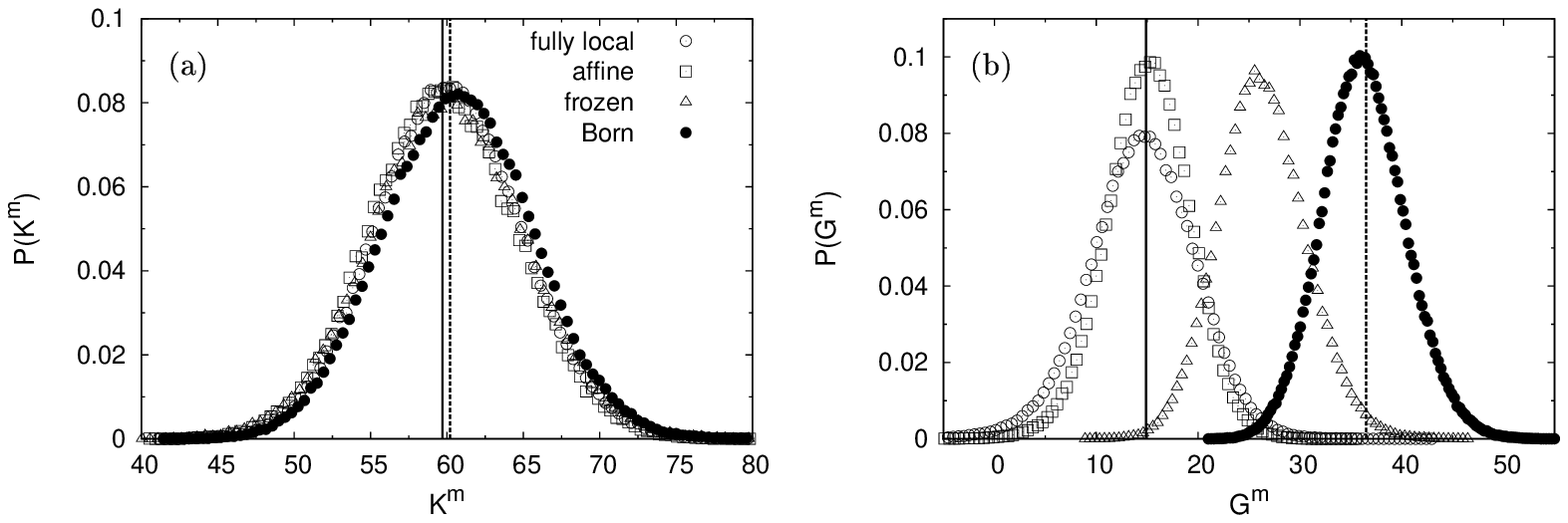}
\vspace*{0mm}
\caption{Comparisons of the probability distributions for (a) bulk modulus and (b) shear modulus for the three approaches: fully local approach (open circle), affine strain approach (square), and frozen matrix approach (triangle).
The local cube size is $W=3.16$. 
The distribution of the Born term, calculated by the fluctuation formula Eq. (\ref{ffmodulus}), is also plotted for comparison (closed circle).
The vertical lines indicate the macroscopic values, $K=59.7$ (solid line) and $K^B=60.2$ (dashed line) in (a), and $G=14.9$ (solid line) and $G^B=36.5$ (dashed line) in (b).
} \label{com_dis}
\end{figure*}

\subsection{Distribution of local elastic moduli}
We next investigated the local moduli quantified by the three approaches described in Sec. \ref{method}.
In Fig. \ref{m_dis}, we show the distributions of the local bulk modulus $K^m$ and the local shear modulus $G^m$.
The considered cube sizes are $W=3.16$, $5.27$, and $7.90$.
We confirmed that the five shear moduli, $G_1^m, G_2^m, G_3^m, G_4^m$, and $G_5^m$ exhibit almost identical distributions, therefore in Fig. \ref{m_dis} we plot data averaged over these five components.
The distributions calculated by the three approaches are all well fitted by   gaussian distributions \cite{tsamados_2009,yoshimoto_2004,makke_2011} (solid lines).
In Fig. \ref{m_dis}, we indicate the values of the macroscopic moduli obtained from Figs. \ref{mbulk} and \ref{mshear} (i.e., $K \simeq 59.7$ and $G \simeq 14.9$) by the vertical lines.
Except for the shear modulus distribution of the frozen matrix approach, all distributions exhibit an average value independent of the cube size $W$, and this average value coincides with the macroscopic one.
The shear modulus distributions obtained in the  frozen matrix approach, in contrast,  exhibits an  average value that depends on the cube size $W$, and  seems to converge to the macroscopic value as $W$ increases, although the convergence  is rather  slow.

In Fig.~\ref{com_dis}, we show the comparison of the distributions obtained from the three approaches for the cube size $W=3.16$.
In the same figure, we also plot the distribution of the Born term (the affine component).
Here it has to be noted that the Born term can be obtained either by the fluctuation formula through $C_{ijkl}^{Bm}$ in Eq. (\ref{ffmodulus}) or by the explicit way, i.e., performing explicit affine deformations as we do in Figs. \ref{mbulk} and \ref{mshear} (``without relaxation" case). We confirmed that these two methods produce identical  distributions of the Born term.
In Fig.~\ref{com_dis} the distribution of the Born term obtained by the fluctuation formula is shown.
From Fig. \ref{com_dis}(a), it is evident that the three approaches produce nearly identical  bulk modulus distributions, and these distributions coincide well with the Born term distribution.
This result indicates that the non-affine component of the local bulk modulus is very small, which leads to the coincidence of the three approaches.
As shown in Fig. \ref{mbulk}, the macroscopic bulk modulus has a very small non-affine component and is mostly determined by the Born term.
The same holds  for the local bulk modulus.

\begin{figure*}[t]
\includegraphics[scale=1, angle=0]{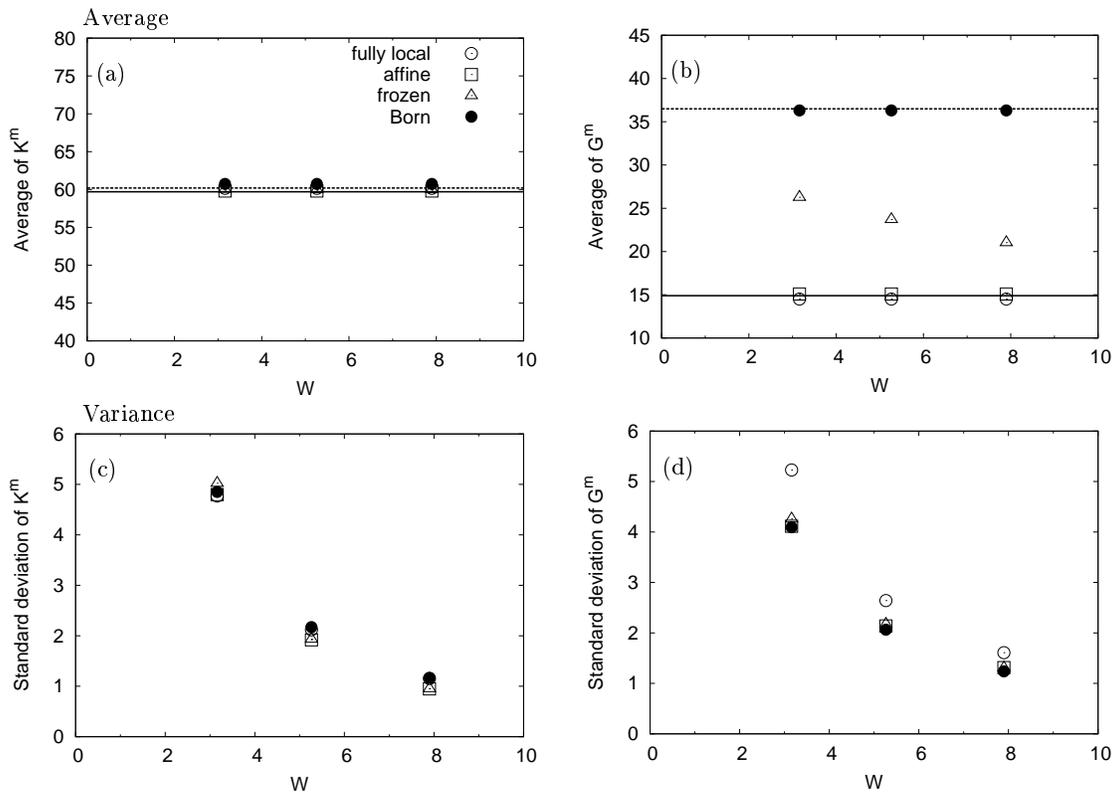}
\vspace*{0mm}
\caption{Comparison of average and standard deviation of the distributions calculated by the three approaches, and of the Born term: fully local approach (open circle), affine strain approach (square), frozen matrix approach (triangle), and the Born term (closed circle). The averages are shown in (a) and (b), and the standard deviations are presented in (c) and (d).
The horizontal lines indicate the macroscopic values, $K=59.7$ (solid line), and $K^B=60.2$ (dashed line) in (a), and $G=14.9$ (solid line) and $G^B=36.5$ (dashed line) in (b).
} \label{com_av}
\end{figure*}

In contrast, the situation is totally different in the case of the local shear modulus.
From Fig. \ref{com_dis}(b), we clearly observe that the shear modulus distribution exhibits large differences among the three approaches. Also, the distributions are very different compared to the Born contribution alone, which means that the local shear modulus has a large non-affine component, as does the macroscopic shear modulus shown in Fig. \ref{mshear}.
The differences among the three approaches are indeed caused by this large non-affine component.
In addition, there are two remarkable differences observed in Fig.~\ref{com_dis}(b).
Firstly, when comparing the distributions calculated by the fully local and the affine strain approaches, we see that the latter  exhibits a narrower distribution (smaller standard deviation). This result indicates that the spatial variations of the local strain field make the shear modulus distribution wider (more heterogeneous).
Secondly, the frozen matrix approach exhibits a much larger average value than the other two methods, whose average values both coincide with the macroscopic one. We believe that the larger average value is caused by the additional constraint resulting from freezing the environment, which limits  the non-affine motions of particles in the cube $m$.
As the cube $m$ can not be fully relaxed during the energy minimization, a larger stress and shear modulus are obtained. To support our explanation, we also observe that the average value of the frozen matrix approach lies between the macroscopic values $G=14.9$ and $G^B=36.5$, which correspond to the values of the non-constrained system and the fully-constrained system, respectively. This result is clearly consistent with the interpretation that the local cube $m$ is only partially relaxed.

In Fig. \ref{com_av}, we compare averages and standard deviations of the distributions calculated by the three approaches for three cube sizes, $W=3.16$, $5.27$, and $7.90$, respectively. From this figure, we can emphasize the differences among the three methods more quantitatively.
For the local bulk modulus, the three approaches exhibit nearly same average and standard deviation values for all three $W$s, and these values coincide well with those estimated from the Born term only.
The average values agree with the macroscopic values $K$ and $K^B$ ($K\simeq K^B$). 
On the other hand, in the case of the local shear modulus, the affine strain approach shows a smaller standard deviation value compared to the fully local one.
Again, this is because the affine strain approach does not consider the spatially varying local strain field.
In addition, the frozen matrix approach exhibits much higher average values than the other two approaches.
The average value of the frozen matrix approach lies between $G$ and $G^B$ and seems to converge to the value $G$ as $W$ gets large. The constraints induced by the  freezing of the matrix prevents the non-affine field from fully contributing to the elastic constants, and causes such large average values.

\begin{figure*}[p!]
\includegraphics[scale=1, angle=0]{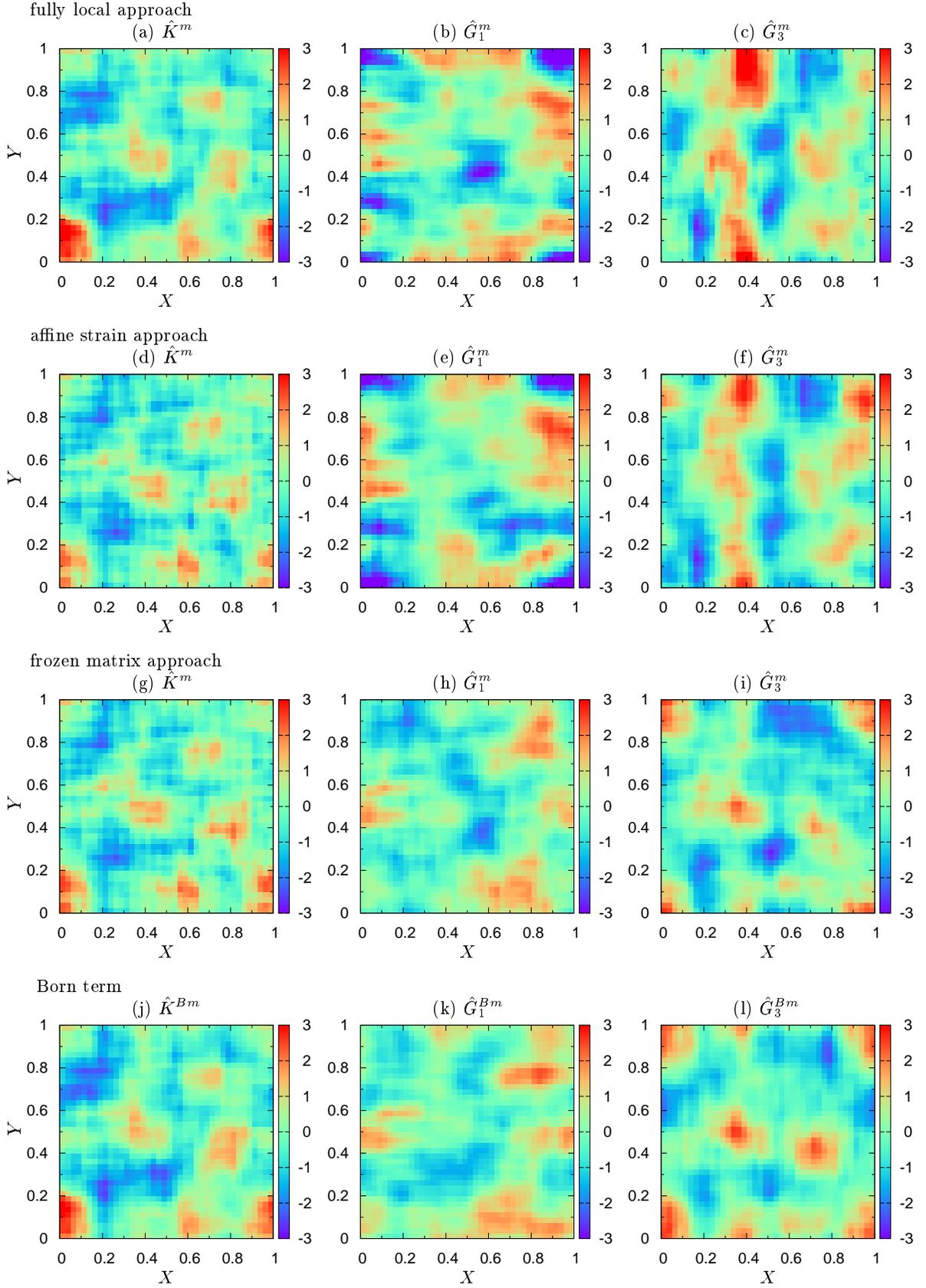}
\vspace*{0mm}
\caption{(Color online)
$xy$-layers taken from the spatial distributions of the bulk modulus $\hat{K}^m$ and the shear moduli $\hat{G}_1^m$, $\hat{G}_3^m$ obtained by three approaches: fully local approach ((a)-(c)), affine strain approach ((d)-(f)), and frozen matrix approach ((g)-(i)).
The value of the local modulus is normalized by its average and standard deviation, i.e., $\hat{X}^m = (X^m-{X}_{\text{ave}})/X_{\text{std}} \ (X \in K,G_{1},G_{3})$ (see Eq. (\ref{nomv})). We also show the spatial maps of the Born terms ((j)-(l)), for comparison.
$X$ and $Y$ coordinates are presented in units of the box length $L=15.8 \sigma$. The $Z$ coordinates of all the layers correspond to $L/2$. Note that no average over the configurations ensemble is involved, data refer to one arbitrary system's instance only.
} \label{com_map}
\end{figure*}

\subsection{Spatial distribution of local elastic modulus}
We have also compared the spatial distributions of the local modulus, represented by color maps,  for the three approaches.
The spatial maps are shown for the case $W=3.16$ in Fig. \ref{com_map}. Here,
we visualize the maps for the variables $\hat{K}^m$ and $\hat{G}^m_i$, which are normalized by both the averages and the standard deviations:
\begin{equation}
\begin{aligned}
\hat{K}^m &= \frac{K^m-K_{\text{ave}}}{K_{\text{std}}}, \\
\hat{G}^m_i &= \frac{G^m_i-G_{i \text{ave}}}{G_{i \text{std}}}, \quad (i \in 1,2,3,4,5), \label{nomv}
\end{aligned}
\end{equation}
where the subscripts ``ave" and ``std" mean average and standard deviation, respectively.
By considering normalized variables $\hat{K}^m$ and $\hat{G}_i^m$, we are able to emphasize domains of the system which are relatively soft or hard. In the same figure we also show the spatial map of the Born term alone, which is also normalized as in Eq. (\ref{nomv}).

\begin{figure*}[t]
\includegraphics[scale=1, angle=0]{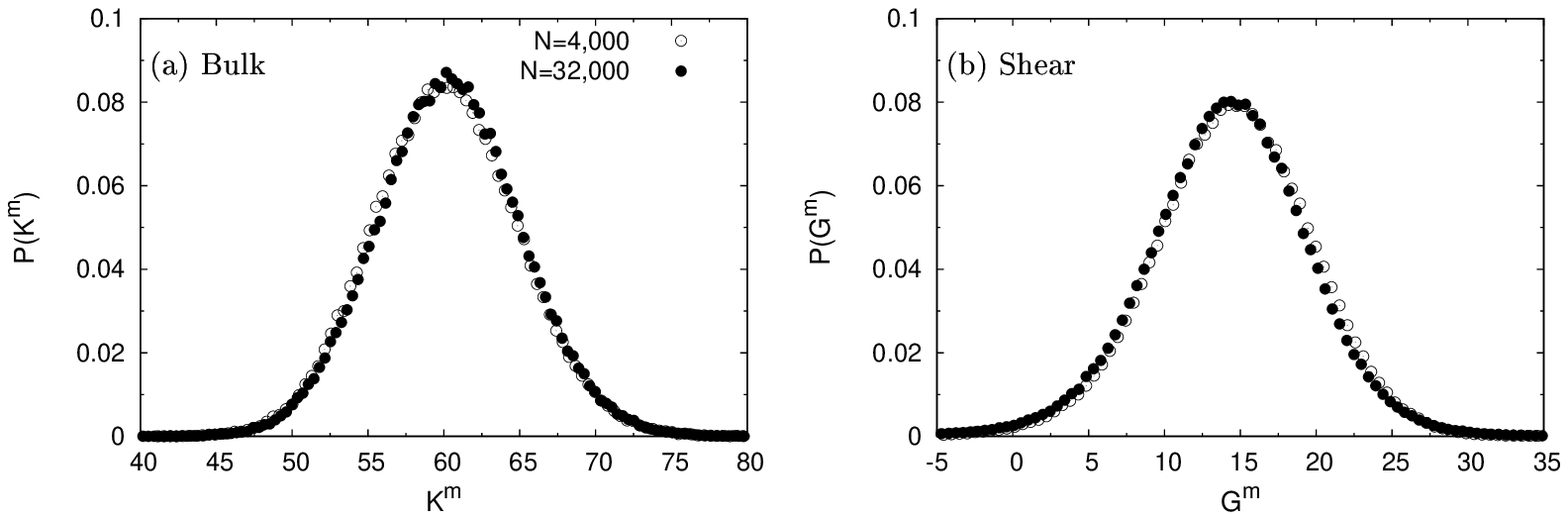}
\vspace*{0mm}
\caption{Comparison of the distributions of (a) bulk modulus and (b) shear modulus for two systems with $N=4,000$ (open circles) and $N=32,000$ (closed circles) particles.
The distributions were calculated by the fluctuation formula Eq. (\ref{ffmodulus}) (i.e., fully local approach).
The local cube size is $W=3.16$. 
} \label{dis_large}
\end{figure*}

For the local bulk modulus $\hat{K}^m$, the spatial maps of the three approaches correlate well with each other and are all very similar to that of the Born term $\hat{K}^{Bm}$.
Again, this result indicates that there is a very small non-affine component in the local bulk modulus, which leads to the very small differences among the three approaches. On the other hand, differences among the three approaches are observed in the maps for the shear moduli $\hat{G}^m_1$ (pure shear) and $\hat{G}^m_3$ (simple shear).
The large non-affine components of the shear moduli indeed change the soft/hard character of different domains in the spatial maps of the Born terms $\hat{G}^{Bm}_1$ and $\hat{G}^{Bm}_3$. The fully local 
approach and the affine strain  exhibit rather similar soft and hard parts, whereas the spatial map of frozen matrix approach looks like that of the Born term, rather than those of the other two approaches.
This observation also indicates that the non-affine component of the frozen matrix approach is limited, due to the constraint induced by the affinely deformed matrix, so that the modifications with respect to the Born term map are limited.

From the above results (comparison of probability distributions and spatial maps), we conclude that all the three approaches can be safely used to measure the local bulk modulus, which is characterized by a small non-affine component.
In contrast, in the case of the local shear modulus, which has a large non-affine component, it is more appropriate to use a fully local approach to deal with both the local stress and the local strain and measure the non-affine component correctly, without any constraint.

We conclude this Section with two remarks.
First,  our results are qualitatively consistent with  previous studies \cite{yoshimoto_2004,makke_2011} on different systems.
The study \cite{yoshimoto_2004} observed that the values of the local shear modulus obtained from the local stress-global strain relation (i.e., affine strain approach) are qualitatively consistent with those obtained from the fluctuation formula Eq. (\ref{ffmodulus}) (i.e.,  fully  local approach).
In our study, the fully local and the affine strain approaches, show similar modulus distributions. Indeed, although the  two approaches exhibit modulus distributions which are different at a quantitative level, the differences are not so large, as evident from Fig. \ref{com_dis}. Moreover, the spatial maps corresponding to the two methods are very similar (see Fig. \ref{com_map}).
This observation is analogous to what noticed in~\cite{yoshimoto_2004}.
In addition, the study \cite{makke_2011} obtained very convincing results by using the frozen matrix approach but only focused 
on the local bulk modulus, ignoring local shear modulus.
Therefore, our present results are consistent with and substantially expand previous similar works \cite{yoshimoto_2004,makke_2011}.

Second, we briefly discuss now the impact of system size on the calculation of local moduli.
Besides the system formed by $N=4,000$ particles which we have used in this study, we also considered a larger system, with $N=32,000$, to check for system size effects. We calculated the local moduli for this larger system by using the fluctuation formula Eq.~(\ref{ffmodulus}) (fully local approach).
We compare the modulus distributions for the two sizes in Fig. \ref{dis_large}.
It is evident that the two systems exhibit the same distributions for both the bulk modulus and the shear modulus.
This result indicates that there are no system size effects on the local modulus.
However, the total length of the MD trajectory used for performing the ensemble average of Eq. (\ref{ffmodulus}) has been found to be very different in the two cases.
While a length  $t=10^5$  was sufficient to obtain converged results for the small system of $N=4,000$, a length of  $t=10^6$ was required for the larger system of $N=32,000$.
More specifically, the correlation term between local and global stress, $\left< \sigma^m_{ij} \sigma_{kl} \right>$, in the non-affine component $C_{ijkl}^{Nm}$, necessitates a longer sampling time  to be estimated correctly. Indeed, the larger system is characterized by long wavelength modes, which contribute to fluctuations of the local stress $\sigma^m_{ij}$ and evolve on slow time scales, which must be correctly sampled. Concluding, convergence with simulation time must be carefully checked for larger  systems.

\section{Conclusions} 
\label{conclusion}
In the present study, we have applied three different approaches, ``fully local", ``affine strain", and ``frozen matrix", to measure the local elastic moduli, bulk modulus and shear modulus of a Lennard-Jones monatomic model glass.
For the case of the local bulk modulus, the three approaches give nearly identical  probability distributions and spatial maps.
This is because the non-affine component in the bulk modulus is relatively small.
The value of the bulk modulus is mostly determined by the Born term (the affine component), and therefore only small differences among the three approaches are observed. However, the situation becomes notably different for the local shear modulus.
The three approaches exhibit different distributions and different spatial maps, even at a qualitative level.
In the case of the shear modulus, the non-affine component has the same order of magnitude as the affine component, which causes the large differences among the three approaches.

The difference between the fully local and the affine strain approaches comes from the use of the local strain or global strain for the calculation of the local modulus. In the affine strain approach, the use of the  global strain instead of the local value results in a standard deviation narrower than that obtained from the fully local approach, where the spatial variations of the local strain field is taken into account. In addition, the difference between the frozen matrix approach and the other two approaches arises from the fact that the system, except for the target local cube, is frozen. In the frozen matrix approach, the constraint of the affine displacement applied to the matrix severely restricts the relaxation in the local region, and the non-affine component of the shear modulus is therefore underestimated. As a result, the average value of the shear modulus is significantly larger than in the other two approaches.
The spatial map of the shear modulus obtained in the frozen matrix approach is rather similar to the Born term map, due to a limited influence of the non-affine component.

Therefore, our conclusion is that one can safely choose among the three approaches to extract correct values for the local bulk modulus, which has a small non-affine component. However, in the case of the local shear modulus, which has a larger non-affine component, only the fully local approach is appropriate, where both local stress and local strain fields are dealt with, and there are no constraints which can limit the correct evolution of the non-affine component. In  this study, we used the fluctuation formula Eq. (\ref{ffmodulus}) to implement the  fully local approach.
An alternative is to obtain the local stress and the local strain relation directly, as was achieved in references~\cite{tsamados_2009,goldhirsch_2010}.
In this case, one needs to measure, in addition to  the local stress field $\sigma_{ij}^m$,
 a local strain field $\epsilon_{ij}^m$, derived from a coarse-grained displacement field.

We conclude with a remark concerning the possible interpretation of the different elastic constants calculated in this paper. 
The computation of local elastic constants can correspond to two different questions. Firstly, one is interested to evaluate the 
local deformation response in a system that is, e.g., subjected to a given macroscopic stress. Clearly, this is
achieved by using the fully local approach (the local fluctuation formula), in which the total deformation of the sample and its influence on the local 
response are properly taken into account. A second objective may be to use the local elastic constants as inputs for a
model at a more coarse grained scale, for example a finite element calculation with heterogeneous elasticity. 
In this case it is clear from the general discussion of the non-affine contributions that the effective elastic constants for the 
bulk system are expected to be {\em smaller} than the average values of the disordered elastic constants used as an input. 
In this perspective, it appears that the use of the frozen matrix approach, which indeed predicts an average value of the local 
modulus that decreases with the coarse graining scale, could be more appropriate than the fully local approach.

 \begin{acknowledgments}
We thank  Prof. P. Sollich and Dr. A. Makke for  helpful correspondence.
The simulations were carried out using LAMMPS molecular dynamics software (http://lammps.sandia.gov)~\cite{plimpton1995fast}. This work was supported by the Nanosciences Foundation of Grenoble. JLB is supported by Institut Universitaire de France. 
\end{acknowledgments}

\bibliographystyle{apsrev4-1}
\bibliography{references}

\begin{thebibliography}{47}%
\makeatletter
\providecommand \@ifxundefined [1]{%
 \@ifx{#1\undefined}
}%
\providecommand \@ifnum [1]{%
 \ifnum #1\expandafter \@firstoftwo
 \else \expandafter \@secondoftwo
 \fi
}%
\providecommand \@ifx [1]{%
 \ifx #1\expandafter \@firstoftwo
 \else \expandafter \@secondoftwo
 \fi
}%
\providecommand \natexlab [1]{#1}%
\providecommand \enquote  [1]{``#1''}%
\providecommand \bibnamefont  [1]{#1}%
\providecommand \bibfnamefont [1]{#1}%
\providecommand \citenamefont [1]{#1}%
\providecommand \href@noop [0]{\@secondoftwo}%
\providecommand \href [0]{\begingroup \@sanitize@url \@href}%
\providecommand \@href[1]{\@@startlink{#1}\@@href}%
\providecommand \@@href[1]{\endgroup#1\@@endlink}%
\providecommand \@sanitize@url [0]{\catcode `\\12\catcode `\$12\catcode
  `\&12\catcode `\#12\catcode `\^12\catcode `\_12\catcode `\%12\relax}%
\providecommand \@@startlink[1]{}%
\providecommand \@@endlink[0]{}%
\providecommand \url  [0]{\begingroup\@sanitize@url \@url }%
\providecommand \@url [1]{\endgroup\@href {#1}{\urlprefix }}%
\providecommand \urlprefix  [0]{URL }%
\providecommand \Eprint [0]{\href }%
\providecommand \doibase [0]{http://dx.doi.org/}%
\providecommand \selectlanguage [0]{\@gobble}%
\providecommand \bibinfo  [0]{\@secondoftwo}%
\providecommand \bibfield  [0]{\@secondoftwo}%
\providecommand \translation [1]{[#1]}%
\providecommand \BibitemOpen [0]{}%
\providecommand \bibitemStop [0]{}%
\providecommand \bibitemNoStop [0]{.\EOS\space}%
\providecommand \EOS [0]{\spacefactor3000\relax}%
\providecommand \BibitemShut  [1]{\csname bibitem#1\endcsname}%
\let\auto@bib@innerbib\@empty
\bibitem [{\citenamefont {Yoshimoto}\ \emph {et~al.}(2004)\citenamefont
  {Yoshimoto}, \citenamefont {Jain}, \citenamefont {Workum}, \citenamefont
  {Nealey},\ and\ \citenamefont {de~Pablo}}]{yoshimoto_2004}%
  \BibitemOpen
  \bibfield  {author} {\bibinfo {author} {\bibfnamefont {K.}~\bibnamefont
  {Yoshimoto}}, \bibinfo {author} {\bibfnamefont {T.~S.}\ \bibnamefont {Jain}},
  \bibinfo {author} {\bibfnamefont {K.~V.}\ \bibnamefont {Workum}}, \bibinfo
  {author} {\bibfnamefont {P.~F.}\ \bibnamefont {Nealey}}, \ and\ \bibinfo
  {author} {\bibfnamefont {J.~J.}\ \bibnamefont {de~Pablo}},\ }\href@noop {}
  {\bibfield  {journal} {\bibinfo  {journal} {Phys. Rev. Lett.}\ }\textbf
  {\bibinfo {volume} {93}},\ \bibinfo {pages} {175501} (\bibinfo {year}
  {2004})}\BibitemShut {NoStop}%
\bibitem [{\citenamefont {Makke}\ \emph {et~al.}(2011)\citenamefont {Makke},
  \citenamefont {Perez}, \citenamefont {Rottler}, \citenamefont {Lame},\ and\
  \citenamefont {Barrat}}]{makke_2011}%
  \BibitemOpen
  \bibfield  {author} {\bibinfo {author} {\bibfnamefont {A.}~\bibnamefont
  {Makke}}, \bibinfo {author} {\bibfnamefont {M.}~\bibnamefont {Perez}},
  \bibinfo {author} {\bibfnamefont {J.}~\bibnamefont {Rottler}}, \bibinfo
  {author} {\bibfnamefont {O.}~\bibnamefont {Lame}}, \ and\ \bibinfo {author}
  {\bibfnamefont {J.-L.}\ \bibnamefont {Barrat}},\ }\href@noop {} {\bibfield
  {journal} {\bibinfo  {journal} {Macromol. Theory Simul.}\ }\textbf {\bibinfo
  {volume} {20}},\ \bibinfo {pages} {826} (\bibinfo {year} {2011})}\BibitemShut
  {NoStop}%
\bibitem [{\citenamefont {Tsamados}\ \emph {et~al.}(2009)\citenamefont
  {Tsamados}, \citenamefont {Tanguy}, \citenamefont {Goldenberg},\ and\
  \citenamefont {Barrat}}]{tsamados_2009}%
  \BibitemOpen
  \bibfield  {author} {\bibinfo {author} {\bibfnamefont {M.}~\bibnamefont
  {Tsamados}}, \bibinfo {author} {\bibfnamefont {A.}~\bibnamefont {Tanguy}},
  \bibinfo {author} {\bibfnamefont {C.}~\bibnamefont {Goldenberg}}, \ and\
  \bibinfo {author} {\bibfnamefont {J.-L.}\ \bibnamefont {Barrat}},\
  }\href@noop {} {\bibfield  {journal} {\bibinfo  {journal} {Phys. Rev. E}\
  }\textbf {\bibinfo {volume} {80}},\ \bibinfo {pages} {026112} (\bibinfo
  {year} {2009})}\BibitemShut {NoStop}%
\bibitem [{\citenamefont {Wittmer}\ \emph {et~al.}(2002)\citenamefont
  {Wittmer}, \citenamefont {Tanguy}, \citenamefont {Barrat},\ and\
  \citenamefont {Lewis}}]{wittmer_2002}%
  \BibitemOpen
  \bibfield  {author} {\bibinfo {author} {\bibfnamefont {J.~P.}\ \bibnamefont
  {Wittmer}}, \bibinfo {author} {\bibfnamefont {A.}~\bibnamefont {Tanguy}},
  \bibinfo {author} {\bibfnamefont {J.-L.}\ \bibnamefont {Barrat}}, \ and\
  \bibinfo {author} {\bibfnamefont {L.}~\bibnamefont {Lewis}},\ }\href@noop {}
  {\bibfield  {journal} {\bibinfo  {journal} {Europhys. Lett.}\ }\textbf
  {\bibinfo {volume} {57}},\ \bibinfo {pages} {423} (\bibinfo {year}
  {2002})}\BibitemShut {NoStop}%
\bibitem [{\citenamefont {Tanguy}\ \emph {et~al.}(2002)\citenamefont {Tanguy},
  \citenamefont {Wittmer}, \citenamefont {Leonforte},\ and\ \citenamefont
  {Barrat}}]{tanguy_2002}%
  \BibitemOpen
  \bibfield  {author} {\bibinfo {author} {\bibfnamefont {A.}~\bibnamefont
  {Tanguy}}, \bibinfo {author} {\bibfnamefont {J.~P.}\ \bibnamefont {Wittmer}},
  \bibinfo {author} {\bibfnamefont {F.}~\bibnamefont {Leonforte}}, \ and\
  \bibinfo {author} {\bibfnamefont {J.-L.}\ \bibnamefont {Barrat}},\
  }\href@noop {} {\bibfield  {journal} {\bibinfo  {journal} {Phys. Rev. B}\
  }\textbf {\bibinfo {volume} {66}},\ \bibinfo {pages} {174205} (\bibinfo
  {year} {2002})}\BibitemShut {NoStop}%
\bibitem [{\citenamefont {Leonforte}\ \emph {et~al.}(2004)\citenamefont
  {Leonforte}, \citenamefont {Tanguy}, \citenamefont {Wittmer},\ and\
  \citenamefont {Barrat}}]{leonforte_2004}%
  \BibitemOpen
  \bibfield  {author} {\bibinfo {author} {\bibfnamefont {F.}~\bibnamefont
  {Leonforte}}, \bibinfo {author} {\bibfnamefont {A.}~\bibnamefont {Tanguy}},
  \bibinfo {author} {\bibfnamefont {J.~P.}\ \bibnamefont {Wittmer}}, \ and\
  \bibinfo {author} {\bibfnamefont {J.-L.}\ \bibnamefont {Barrat}},\
  }\href@noop {} {\bibfield  {journal} {\bibinfo  {journal} {Phys. Rev. B}\
  }\textbf {\bibinfo {volume} {70}},\ \bibinfo {pages} {014203} (\bibinfo
  {year} {2004})}\BibitemShut {NoStop}%
\bibitem [{\citenamefont {Leonforte}\ \emph {et~al.}(2005)\citenamefont
  {Leonforte}, \citenamefont {Boissi$\grave{\text{e}}$re}, \citenamefont
  {Tanguy}, \citenamefont {Wittmer},\ and\ \citenamefont
  {Barrat}}]{leonforte_2005}%
  \BibitemOpen
  \bibfield  {author} {\bibinfo {author} {\bibfnamefont {F.}~\bibnamefont
  {Leonforte}}, \bibinfo {author} {\bibfnamefont {R.}~\bibnamefont
  {Boissi$\grave{\text{e}}$re}}, \bibinfo {author} {\bibfnamefont
  {A.}~\bibnamefont {Tanguy}}, \bibinfo {author} {\bibfnamefont {J.~P.}\
  \bibnamefont {Wittmer}}, \ and\ \bibinfo {author} {\bibfnamefont {J.-L.}\
  \bibnamefont {Barrat}},\ }\href@noop {} {\bibfield  {journal} {\bibinfo
  {journal} {Phys. Rev. B}\ }\textbf {\bibinfo {volume} {72}},\ \bibinfo
  {pages} {224206} (\bibinfo {year} {2005})}\BibitemShut {NoStop}%
\bibitem [{\citenamefont {DiDonna}\ and\ \citenamefont
  {Lubensky}(2005)}]{Didonna}%
  \BibitemOpen
  \bibfield  {author} {\bibinfo {author} {\bibfnamefont {B.~A.}\ \bibnamefont
  {DiDonna}}\ and\ \bibinfo {author} {\bibfnamefont {T.~C.}\ \bibnamefont
  {Lubensky}},\ }\href@noop {} {\bibfield  {journal} {\bibinfo  {journal}
  {Phys. Rev. E}\ }\textbf {\bibinfo {volume} {72}},\ \bibinfo {pages} {066619}
  (\bibinfo {year} {2005})}\BibitemShut {NoStop}%
\bibitem [{\citenamefont {Maloney}(2006)}]{Maloney}%
  \BibitemOpen
  \bibfield  {author} {\bibinfo {author} {\bibfnamefont {C.~E.}\ \bibnamefont
  {Maloney}},\ }\href@noop {} {\bibfield  {journal} {\bibinfo  {journal} {Phys.
  Rev. Lett.}\ }\textbf {\bibinfo {volume} {97}},\ \bibinfo {pages} {035503}
  (\bibinfo {year} {2006})}\BibitemShut {NoStop}%
\bibitem [{\citenamefont {Monaco}\ and\ \citenamefont
  {Giordano}(2009)}]{monaco_2009}%
  \BibitemOpen
  \bibfield  {author} {\bibinfo {author} {\bibfnamefont {G.}~\bibnamefont
  {Monaco}}\ and\ \bibinfo {author} {\bibfnamefont {V.~M.}\ \bibnamefont
  {Giordano}},\ }\href@noop {} {\bibfield  {journal} {\bibinfo  {journal}
  {Proc. Natl. Acad. Sci. USA}\ }\textbf {\bibinfo {volume} {106}},\ \bibinfo
  {pages} {3659} (\bibinfo {year} {2009})}\BibitemShut {NoStop}%
\bibitem [{\citenamefont {Monaco}\ and\ \citenamefont
  {Mossa}(2009)}]{monaco2_2009}%
  \BibitemOpen
  \bibfield  {author} {\bibinfo {author} {\bibfnamefont {G.}~\bibnamefont
  {Monaco}}\ and\ \bibinfo {author} {\bibfnamefont {S.}~\bibnamefont {Mossa}},\
  }\href@noop {} {\bibfield  {journal} {\bibinfo  {journal} {Proc. Natl. Acad.
  Sci. USA}\ }\textbf {\bibinfo {volume} {106}},\ \bibinfo {pages} {16907}
  (\bibinfo {year} {2009})}\BibitemShut {NoStop}%
\bibitem [{\citenamefont {Phillips}(1981)}]{lowtem}%
  \BibitemOpen
  \bibfield  {author} {\bibinfo {author} {\bibfnamefont {W.~A.}\ \bibnamefont
  {Phillips}},\ }\href@noop {} {\emph {\bibinfo {title} {Amorphous Solids: Low
  Temperature Properties}}},\ \bibinfo {edition} {3rd}\ ed.\ (\bibinfo
  {publisher} {Springer, Berlin},\ \bibinfo {year} {1981})\BibitemShut
  {NoStop}%
\bibitem [{\citenamefont {Buchenau}\ \emph {et~al.}(1984)\citenamefont
  {Buchenau}, \citenamefont {N$\ddot{\text{u}}$cker},\ and\ \citenamefont
  {Dianoux}}]{buchenau_1984}%
  \BibitemOpen
  \bibfield  {author} {\bibinfo {author} {\bibfnamefont {U.}~\bibnamefont
  {Buchenau}}, \bibinfo {author} {\bibfnamefont {N.}~\bibnamefont
  {N$\ddot{\text{u}}$cker}}, \ and\ \bibinfo {author} {\bibfnamefont {A.~J.}\
  \bibnamefont {Dianoux}},\ }\href@noop {} {\bibfield  {journal} {\bibinfo
  {journal} {Phys. Rev. Lett.}\ }\textbf {\bibinfo {volume} {53}},\ \bibinfo
  {pages} {2316} (\bibinfo {year} {1984})}\BibitemShut {NoStop}%
\bibitem [{\citenamefont {Malinovsky}\ \emph {et~al.}(1991)\citenamefont
  {Malinovsky}, \citenamefont {Novikov},\ and\ \citenamefont
  {Sokolov}}]{malinovsky_1991}%
  \BibitemOpen
  \bibfield  {author} {\bibinfo {author} {\bibfnamefont {V.}~\bibnamefont
  {Malinovsky}}, \bibinfo {author} {\bibfnamefont {V.~N.}\ \bibnamefont
  {Novikov}}, \ and\ \bibinfo {author} {\bibfnamefont {A.~P.}\ \bibnamefont
  {Sokolov}},\ }\href@noop {} {\bibfield  {journal} {\bibinfo  {journal} {Phys.
  Lett. A}\ }\textbf {\bibinfo {volume} {153}},\ \bibinfo {pages} {63}
  (\bibinfo {year} {1991})}\BibitemShut {NoStop}%
\bibitem [{\citenamefont {Ruffl$\acute{\text{e}}$}\ \emph
  {et~al.}(2003)\citenamefont {Ruffl$\acute{\text{e}}$}, \citenamefont {Foret},
  \citenamefont {Courtens}, \citenamefont {Vacher},\ and\ \citenamefont
  {Monaco}}]{ruffle_2003}%
  \BibitemOpen
  \bibfield  {author} {\bibinfo {author} {\bibfnamefont {B.}~\bibnamefont
  {Ruffl$\acute{\text{e}}$}}, \bibinfo {author} {\bibfnamefont
  {M.}~\bibnamefont {Foret}}, \bibinfo {author} {\bibfnamefont
  {E.}~\bibnamefont {Courtens}}, \bibinfo {author} {\bibfnamefont
  {R.}~\bibnamefont {Vacher}}, \ and\ \bibinfo {author} {\bibfnamefont
  {G.}~\bibnamefont {Monaco}},\ }\href@noop {} {\bibfield  {journal} {\bibinfo
  {journal} {Phys. Rev. Lett.}\ }\textbf {\bibinfo {volume} {90}},\ \bibinfo
  {pages} {095502} (\bibinfo {year} {2003})}\BibitemShut {NoStop}%
\bibitem [{\citenamefont {Ruffl$\acute{\text{e}}$}\ \emph
  {et~al.}(2006)\citenamefont {Ruffl$\acute{\text{e}}$}, \citenamefont
  {Guimbreti$\grave{\text{e}}$re}, \citenamefont {Courtens}, \citenamefont
  {Vacher},\ and\ \citenamefont {Monaco}}]{ruffle_2006}%
  \BibitemOpen
  \bibfield  {author} {\bibinfo {author} {\bibfnamefont {B.}~\bibnamefont
  {Ruffl$\acute{\text{e}}$}}, \bibinfo {author} {\bibfnamefont
  {G.}~\bibnamefont {Guimbreti$\grave{\text{e}}$re}}, \bibinfo {author}
  {\bibfnamefont {E.}~\bibnamefont {Courtens}}, \bibinfo {author}
  {\bibfnamefont {R.}~\bibnamefont {Vacher}}, \ and\ \bibinfo {author}
  {\bibfnamefont {G.}~\bibnamefont {Monaco}},\ }\href@noop {} {\bibfield
  {journal} {\bibinfo  {journal} {Phys. Rev. Lett.}\ }\textbf {\bibinfo
  {volume} {96}},\ \bibinfo {pages} {045502} (\bibinfo {year}
  {2006})}\BibitemShut {NoStop}%
\bibitem [{\citenamefont {Masciovecchio}\ \emph {et~al.}(2006)\citenamefont
  {Masciovecchio}, \citenamefont {Baldi}, \citenamefont {Caponi}, \citenamefont
  {Comez}, \citenamefont {Fonzo}, \citenamefont {Fioretto}, \citenamefont
  {Fontana}, \citenamefont {Gessini}, \citenamefont {Santucci}, \citenamefont
  {Sette}, \citenamefont {Viliani}, \citenamefont {Vilmercati},\ and\
  \citenamefont {Ruocco}}]{masciovecchio_2006}%
  \BibitemOpen
  \bibfield  {author} {\bibinfo {author} {\bibfnamefont {C.}~\bibnamefont
  {Masciovecchio}}, \bibinfo {author} {\bibfnamefont {G.}~\bibnamefont
  {Baldi}}, \bibinfo {author} {\bibfnamefont {S.}~\bibnamefont {Caponi}},
  \bibinfo {author} {\bibfnamefont {L.}~\bibnamefont {Comez}}, \bibinfo
  {author} {\bibfnamefont {S.~D.}\ \bibnamefont {Fonzo}}, \bibinfo {author}
  {\bibfnamefont {D.}~\bibnamefont {Fioretto}}, \bibinfo {author}
  {\bibfnamefont {A.}~\bibnamefont {Fontana}}, \bibinfo {author} {\bibfnamefont
  {A.}~\bibnamefont {Gessini}}, \bibinfo {author} {\bibfnamefont {S.~C.}\
  \bibnamefont {Santucci}}, \bibinfo {author} {\bibfnamefont {F.}~\bibnamefont
  {Sette}}, \bibinfo {author} {\bibfnamefont {G.}~\bibnamefont {Viliani}},
  \bibinfo {author} {\bibfnamefont {P.}~\bibnamefont {Vilmercati}}, \ and\
  \bibinfo {author} {\bibfnamefont {G.}~\bibnamefont {Ruocco}},\ }\href@noop {}
  {\bibfield  {journal} {\bibinfo  {journal} {Phys. Rev. Lett.}\ }\textbf
  {\bibinfo {volume} {97}},\ \bibinfo {pages} {035501} (\bibinfo {year}
  {2006})}\BibitemShut {NoStop}%
\bibitem [{\citenamefont {Schirmacher}(2006)}]{schirmacher_2006}%
  \BibitemOpen
  \bibfield  {author} {\bibinfo {author} {\bibfnamefont {W.}~\bibnamefont
  {Schirmacher}},\ }\href@noop {} {\bibfield  {journal} {\bibinfo  {journal}
  {Europhys. Lett.}\ }\textbf {\bibinfo {volume} {73}},\ \bibinfo {pages} {892}
  (\bibinfo {year} {2006})}\BibitemShut {NoStop}%
\bibitem [{\citenamefont {Schirmacher}\ \emph {et~al.}(2007)\citenamefont
  {Schirmacher}, \citenamefont {Ruocco},\ and\ \citenamefont
  {Scopigno}}]{schirmacher_2007}%
  \BibitemOpen
  \bibfield  {author} {\bibinfo {author} {\bibfnamefont {W.}~\bibnamefont
  {Schirmacher}}, \bibinfo {author} {\bibfnamefont {G.}~\bibnamefont {Ruocco}},
  \ and\ \bibinfo {author} {\bibfnamefont {T.}~\bibnamefont {Scopigno}},\
  }\href@noop {} {\bibfield  {journal} {\bibinfo  {journal} {Phys. Rev. Lett.}\
  }\textbf {\bibinfo {volume} {98}},\ \bibinfo {pages} {025501} (\bibinfo
  {year} {2007})}\BibitemShut {NoStop}%
\bibitem [{\citenamefont {Tanguy}\ \emph {et~al.}(2006)\citenamefont {Tanguy},
  \citenamefont {Leonforte},\ and\ \citenamefont {Barrat}}]{tanguy_2006}%
  \BibitemOpen
  \bibfield  {author} {\bibinfo {author} {\bibfnamefont {A.}~\bibnamefont
  {Tanguy}}, \bibinfo {author} {\bibfnamefont {F.}~\bibnamefont {Leonforte}}, \
  and\ \bibinfo {author} {\bibfnamefont {J.-L.}\ \bibnamefont {Barrat}},\
  }\href@noop {} {\bibfield  {journal} {\bibinfo  {journal} {Eur. Phys. J. E}\
  }\textbf {\bibinfo {volume} {20}},\ \bibinfo {pages} {355} (\bibinfo {year}
  {2006})}\BibitemShut {NoStop}%
\bibitem [{\citenamefont {Tanguy}\ \emph {et~al.}(2010)\citenamefont {Tanguy},
  \citenamefont {Mantisi},\ and\ \citenamefont {Tsamados}}]{tanguy_2010}%
  \BibitemOpen
  \bibfield  {author} {\bibinfo {author} {\bibfnamefont {A.}~\bibnamefont
  {Tanguy}}, \bibinfo {author} {\bibfnamefont {B.}~\bibnamefont {Mantisi}}, \
  and\ \bibinfo {author} {\bibfnamefont {M.}~\bibnamefont {Tsamados}},\
  }\href@noop {} {\bibfield  {journal} {\bibinfo  {journal} {Europhys. Lett.}\
  }\textbf {\bibinfo {volume} {90}},\ \bibinfo {pages} {16004} (\bibinfo {year}
  {2010})}\BibitemShut {NoStop}%
\bibitem [{\citenamefont {Lutsko}(1988)}]{lutsko_1988}%
  \BibitemOpen
  \bibfield  {author} {\bibinfo {author} {\bibfnamefont {J.~F.}\ \bibnamefont
  {Lutsko}},\ }\href@noop {} {\bibfield  {journal} {\bibinfo  {journal} {J.
  Appl. Phys.}\ }\textbf {\bibinfo {volume} {64}},\ \bibinfo {pages} {1152}
  (\bibinfo {year} {1988})}\BibitemShut {NoStop}%
\bibitem [{\citenamefont {Cormier}\ \emph {et~al.}(2001)\citenamefont
  {Cormier}, \citenamefont {Rickman},\ and\ \citenamefont
  {Delph}}]{cormier_2001}%
  \BibitemOpen
  \bibfield  {author} {\bibinfo {author} {\bibfnamefont {J.}~\bibnamefont
  {Cormier}}, \bibinfo {author} {\bibfnamefont {J.~M.}\ \bibnamefont
  {Rickman}}, \ and\ \bibinfo {author} {\bibfnamefont {T.~J.}\ \bibnamefont
  {Delph}},\ }\href@noop {} {\bibfield  {journal} {\bibinfo  {journal} {J.
  Appl. Phys.}\ }\textbf {\bibinfo {volume} {89}},\ \bibinfo {pages} {99}
  (\bibinfo {year} {2001})}\BibitemShut {NoStop}%
\bibitem [{\citenamefont {Goldhirsch}\ and\ \citenamefont
  {Goldenberg}(2002)}]{goldhirsch_2010}%
  \BibitemOpen
  \bibfield  {author} {\bibinfo {author} {\bibfnamefont {I.}~\bibnamefont
  {Goldhirsch}}\ and\ \bibinfo {author} {\bibfnamefont {C.}~\bibnamefont
  {Goldenberg}},\ }\href@noop {} {\bibfield  {journal} {\bibinfo  {journal}
  {Eur. Phys. J. E}\ }\textbf {\bibinfo {volume} {9}},\ \bibinfo {pages} {245}
  (\bibinfo {year} {2002})}\BibitemShut {NoStop}%
\bibitem [{\citenamefont {Sollich}()}]{sollich_2009}%
  \BibitemOpen
  \bibfield  {author} {\bibinfo {author} {\bibfnamefont {P.}~\bibnamefont
  {Sollich}},\ }\href@noop {} {\ }\bibinfo {note} {Lecture notes, School on
  Glass Formers and Glasses (Jan. 4-20, 2010), Bangalore, India}\BibitemShut
  {NoStop}%
\bibitem [{\citenamefont {Sollich}\ and\ \citenamefont
  {Barra}()}]{sollich_2012}%
  \BibitemOpen
  \bibfield  {author} {\bibinfo {author} {\bibfnamefont {P.}~\bibnamefont
  {Sollich}}\ and\ \bibinfo {author} {\bibfnamefont {A.}~\bibnamefont
  {Barra}},\ }\href@noop {} {\ }\bibinfo {note} {In preparation}\BibitemShut
  {NoStop}%
\bibitem [{\citenamefont {Lutsko}(1989)}]{lutsko_1989}%
  \BibitemOpen
  \bibfield  {author} {\bibinfo {author} {\bibfnamefont {J.~F.}\ \bibnamefont
  {Lutsko}},\ }\href@noop {} {\bibfield  {journal} {\bibinfo  {journal} {J.
  Appl. Phys.}\ }\textbf {\bibinfo {volume} {65}},\ \bibinfo {pages} {2991}
  (\bibinfo {year} {1989})}\BibitemShut {NoStop}%
\bibitem [{\citenamefont {Wittmer}\ \emph {et~al.}(2013)\citenamefont
  {Wittmer}, \citenamefont {Xu}, \citenamefont {Poli{\'n}ska}, \citenamefont
  {Weysser},\ and\ \citenamefont {Baschnagel}}]{wittmer2013shear}%
  \BibitemOpen
  \bibfield  {author} {\bibinfo {author} {\bibfnamefont {J.}~\bibnamefont
  {Wittmer}}, \bibinfo {author} {\bibfnamefont {H.}~\bibnamefont {Xu}},
  \bibinfo {author} {\bibfnamefont {P.}~\bibnamefont {Poli{\'n}ska}}, \bibinfo
  {author} {\bibfnamefont {F.}~\bibnamefont {Weysser}}, \ and\ \bibinfo
  {author} {\bibfnamefont {J.}~\bibnamefont {Baschnagel}},\ }\href@noop {}
  {\bibfield  {journal} {\bibinfo  {journal} {J. Chem. Phys.}\ }\textbf
  {\bibinfo {volume} {138}},\ \bibinfo {pages} {12A533} (\bibinfo {year}
  {2013})}\BibitemShut {NoStop}%
\bibitem [{\citenamefont {Barron}\ and\ \citenamefont
  {Klein}(1965)}]{barron_1965}%
  \BibitemOpen
  \bibfield  {author} {\bibinfo {author} {\bibfnamefont {T.~H.~K.}\
  \bibnamefont {Barron}}\ and\ \bibinfo {author} {\bibfnamefont {M.~L.}\
  \bibnamefont {Klein}},\ }\href@noop {} {\bibfield  {journal} {\bibinfo
  {journal} {Proc. Phys. Soc. London}\ }\textbf {\bibinfo {volume} {85}},\
  \bibinfo {pages} {523} (\bibinfo {year} {1965})}\BibitemShut {NoStop}%
\bibitem [{Note1()}]{Note1}%
  \BibitemOpen
  \bibinfo {note} {By considering the existence of a strain-energy function, we
  obtain the symmetry of the macroscopic modulus $C_{ijkl}$,
  $C_{ijkl}-C_{klij}=\sigma ^{0}_{kl}\delta _{ij}-\sigma ^{0}_{ij}\delta
  _{kl}$, where $\sigma ^{0}_{ij}$ is the macroscopic initial stress, and
  $\delta _{ij}$ is the Kronecker delta (unity when $i=j$, zero otherwise) (see
  Eq. (4.20) in Ref. \cite {barron_1965}). When $\sigma ^{0}_{ij}$ is
  isotropic, which is often true at macroscopic scale, then
  $C_{ijkl}=C_{klij}$. However, such a strain-energy function does not
  necessarily exist at local scale.}\BibitemShut {Stop}%
\bibitem [{\citenamefont {Fung}(1977)}]{continuum}%
  \BibitemOpen
  \bibfield  {author} {\bibinfo {author} {\bibfnamefont {Y.~C.}\ \bibnamefont
  {Fung}},\ }\href@noop {} {\emph {\bibinfo {title} {A First Course In
  Continuum Mechanics}}},\ \bibinfo {edition} {2nd}\ ed.\ (\bibinfo
  {publisher} {Prentice-Hall, New Jersey, U. S. A},\ \bibinfo {year}
  {1977})\BibitemShut {NoStop}%
\bibitem [{\citenamefont {Barrat}\ \emph {et~al.}(1988)\citenamefont {Barrat},
  \citenamefont {Roux}, \citenamefont {Hansen},\ and\ \citenamefont
  {Klein}}]{barrat_1988}%
  \BibitemOpen
  \bibfield  {author} {\bibinfo {author} {\bibfnamefont {J.-L.}\ \bibnamefont
  {Barrat}}, \bibinfo {author} {\bibfnamefont {J.-N.}\ \bibnamefont {Roux}},
  \bibinfo {author} {\bibfnamefont {J.-P.}\ \bibnamefont {Hansen}}, \ and\
  \bibinfo {author} {\bibfnamefont {M.~L.}\ \bibnamefont {Klein}},\ }\href@noop
  {} {\bibfield  {journal} {\bibinfo  {journal} {Europhys. Lett.}\ }\textbf
  {\bibinfo {volume} {7}},\ \bibinfo {pages} {707} (\bibinfo {year}
  {1988})}\BibitemShut {NoStop}%
\bibitem [{Note2()}]{Note2}%
  \BibitemOpen
  \bibinfo {note} {The Born term $C_{ijkl}^{Bm}$ included in the calculation of
  the modulus tensor in Eq. (\ref {ffmodulus}), involves the second derivative
  of the potential. Due to the truncation of the potential, this gives an
  impulsive correction to the Born term~\cite {Xu2012impulsive}. We have
  checked that this correction is always negligible in our system.}\BibitemShut
  {Stop}%
\bibitem [{\citenamefont {Robles}\ and\ \citenamefont
  {de~Haro}(2003)}]{robles_2003}%
  \BibitemOpen
  \bibfield  {author} {\bibinfo {author} {\bibfnamefont {M.}~\bibnamefont
  {Robles}}\ and\ \bibinfo {author} {\bibfnamefont {M.~L.}\ \bibnamefont
  {de~Haro}},\ }\href@noop {} {\bibfield  {journal} {\bibinfo  {journal}
  {Europhys. Lett.}\ }\textbf {\bibinfo {volume} {62}},\ \bibinfo {pages} {56}
  (\bibinfo {year} {2003})}\BibitemShut {NoStop}%
\bibitem [{\citenamefont {Squire}\ \emph {et~al.}(1968)\citenamefont {Squire},
  \citenamefont {Holt},\ and\ \citenamefont {Hoover}}]{squirey_1968}%
  \BibitemOpen
  \bibfield  {author} {\bibinfo {author} {\bibfnamefont {D.~R.}\ \bibnamefont
  {Squire}}, \bibinfo {author} {\bibfnamefont {A.~C.}\ \bibnamefont {Holt}}, \
  and\ \bibinfo {author} {\bibfnamefont {W.~G.}\ \bibnamefont {Hoover}},\
  }\href@noop {} {\bibfield  {journal} {\bibinfo  {journal} {Physica}\ }\textbf
  {\bibinfo {volume} {42}},\ \bibinfo {pages} {388} (\bibinfo {year}
  {1968})}\BibitemShut {NoStop}%
\bibitem [{\citenamefont {Ray}\ and\ \citenamefont {Rahman}(1984)}]{ray_1984}%
  \BibitemOpen
  \bibfield  {author} {\bibinfo {author} {\bibfnamefont {J.~R.}\ \bibnamefont
  {Ray}}\ and\ \bibinfo {author} {\bibfnamefont {A.}~\bibnamefont {Rahman}},\
  }\href@noop {} {\bibfield  {journal} {\bibinfo  {journal} {J. Chem. Phys.}\
  }\textbf {\bibinfo {volume} {80}},\ \bibinfo {pages} {4423} (\bibinfo {year}
  {1984})}\BibitemShut {NoStop}%
\bibitem [{\citenamefont {Ray}\ \emph {et~al.}(1985)\citenamefont {Ray},
  \citenamefont {Moody},\ and\ \citenamefont {Rahman}}]{ray_1985}%
  \BibitemOpen
  \bibfield  {author} {\bibinfo {author} {\bibfnamefont {J.~R.}\ \bibnamefont
  {Ray}}, \bibinfo {author} {\bibfnamefont {M.~C.}\ \bibnamefont {Moody}}, \
  and\ \bibinfo {author} {\bibfnamefont {A.}~\bibnamefont {Rahman}},\
  }\href@noop {} {\bibfield  {journal} {\bibinfo  {journal} {Phys. Rev. B}\
  }\textbf {\bibinfo {volume} {32}},\ \bibinfo {pages} {733} (\bibinfo {year}
  {1985})}\BibitemShut {NoStop}%
\bibitem [{\citenamefont {Hansen}\ and\ \citenamefont
  {McDonald}(2006)}]{simpleliquid}%
  \BibitemOpen
  \bibfield  {author} {\bibinfo {author} {\bibfnamefont {J.~P.}\ \bibnamefont
  {Hansen}}\ and\ \bibinfo {author} {\bibfnamefont {I.~R.}\ \bibnamefont
  {McDonald}},\ }\href@noop {} {\emph {\bibinfo {title} {Theory of Simple
  Liquids}}},\ \bibinfo {edition} {3rd}\ ed.\ (\bibinfo  {publisher} {Academic,
  London},\ \bibinfo {year} {2006})\BibitemShut {NoStop}%
\bibitem [{Note3()}]{Note3}%
  \BibitemOpen
  \bibinfo {note} {The Eq. (\ref {ffmodulus}) (fluctuation formula) is
  formulated from the second derivative of the local energy density with
  respect to the local Green-Lagrange strain tensor \cite
  {lutsko_1988,yoshimoto_2004}. When the initial stress $\unhbox \voidb@x \hbox
  {\protect \boldmath $\sigma $}^{0m}$ has finite values, a deviation from the
  definition of Eq. (\ref {lssr}) (the first derivative of the stress with
  respect to the strain) appears. This deviation is taken care of by adding the
  correction term $C_{ijkl}^{Cm}$ to the modulus $C_{ijkl}^m$ in Eq.~(\ref
  {ffmodulus}) (see Eq.~(4.19) in Ref.~\cite {barron_1965}): \begin {equation}
  \begin {aligned} C_{ijkl}^{Cm} & = -\protect \frac {1}{2} {\setbox \z@ \hbox
  {\frozen@everymath \@emptytoks \mathsurround \z@ $\nulldelimiterspace \z@
  \left (\vcenter to\@ne \big@size {}\right .$}\box \z@ } 2\left <\sigma
  ^m_{ij}\right >\delta _{kl}-\left <\sigma ^m_{ik}\right >\delta _{jl}-\left
  <\sigma ^m_{il}\right >\delta _{jk} \\ & \hskip 2em\relax \hskip 1em\relax
  -\left <\sigma ^m_{jk}\right >\delta _{il}-\left <\sigma ^m_{jl}\right
  >\delta _{ik} {\setbox \z@ \hbox {\frozen@everymath \@emptytoks \mathsurround
  \z@ $\nulldelimiterspace \z@ \left )\vcenter to\@ne \big@size {}\right
  .$}\box \z@ }. \end {aligned} \label {ffcorrect} \end {equation} In this
  study we included the correction term $C_{ijkl}^{Cm}$ into the Born term
  $C_{ijkl}^{Bm}$. We have checked that the present system has small values of
  the initial stress $\unhbox \voidb@x \hbox {\protect \boldmath $\sigma
  $}^{0m}$, leading to a small contribution $C_{ijkl}^{Cm}$.}\BibitemShut
  {Stop}%
\bibitem [{Note4()}]{Note4}%
  \BibitemOpen
  \bibinfo {note} {By substituting $\sigma _{ij}=(1/V)\DOTSB \sum@ \slimits@ _m
  W^3 \sigma ^m_{ij}$ in $C_{ijkl}^{Nm}$ of Eq.~(\ref {ffmodulus}), the
  non-affine component $C_{ijkl}^{Nm}$ is written as \begin {equation}
  C_{ijkl}^{Nm} = \DOTSB \sum@ \slimits@ _{n} \protect \frac {W^3}{T} [\left <
  \sigma ^m_{ij} \sigma ^n_{kl} \right >-\left < \sigma ^m_{ij} \right >\left <
  \sigma ^n_{kl} \right >], \end {equation} from which we can consider
  $C_{ijkl}^{Nm}$ as the sum of correlations of local stress
  fluctuations.}\BibitemShut {Stop}%
\bibitem [{\citenamefont {Irving}\ and\ \citenamefont
  {Kirkwood}(1950)}]{irving_1950}%
  \BibitemOpen
  \bibfield  {author} {\bibinfo {author} {\bibfnamefont {J.~H.}\ \bibnamefont
  {Irving}}\ and\ \bibinfo {author} {\bibfnamefont {J.~G.}\ \bibnamefont
  {Kirkwood}},\ }\href@noop {} {\bibfield  {journal} {\bibinfo  {journal} {J.
  Chem. Phys.}\ }\textbf {\bibinfo {volume} {18}},\ \bibinfo {pages} {817}
  (\bibinfo {year} {1950})}\BibitemShut {NoStop}%
\bibitem [{\citenamefont {MacNeill}\ and\ \citenamefont
  {Rottler}(2010)}]{macneill_2010}%
  \BibitemOpen
  \bibfield  {author} {\bibinfo {author} {\bibfnamefont {D.}~\bibnamefont
  {MacNeill}}\ and\ \bibinfo {author} {\bibfnamefont {J.}~\bibnamefont
  {Rottler}},\ }\href@noop {} {\bibfield  {journal} {\bibinfo  {journal} {Phys.
  Rev. E}\ }\textbf {\bibinfo {volume} {81}},\ \bibinfo {pages} {011804}
  (\bibinfo {year} {2010})}\BibitemShut {NoStop}%
\bibitem [{Note5()}]{Note5}%
  \BibitemOpen
  \bibinfo {note} {The exact statistical mechanics derivation allows one to
  establish Eq.(\ref {ffstress}) for the local stress tensor, which exactly
  conserves momentum. Note that here the quantity $q^{ab}$ is exactly the
  length of the line segment $r_i^{ab}$ located inside the cube. In particular,
  it also accounts for the cases where both $a$ and $b$ are outside the cube.
  In contrast, Eq.(\ref {gsstress}) equally shares the contribution to the
  summation between the two atoms. We have chosen to consider this simplified
  formulation for computational purposes and checked that corrections to the
  exact form are negligible.}\BibitemShut {Stop}%
\bibitem [{\citenamefont {Allen}\ and\ \citenamefont
  {Tildesley}(1986)}]{Allen1986}%
  \BibitemOpen
  \bibfield  {author} {\bibinfo {author} {\bibfnamefont {M.~P.}\ \bibnamefont
  {Allen}}\ and\ \bibinfo {author} {\bibfnamefont {D.~J.}\ \bibnamefont
  {Tildesley}},\ }\href@noop {} {\emph {\bibinfo {title} {Computer Simulation
  of Liquids}}}\ (\bibinfo  {publisher} {Oxford University Press, Oxford},\
  \bibinfo {year} {1986})\BibitemShut {NoStop}%
\bibitem [{Note6()}]{Note6}%
  \BibitemOpen
  \bibinfo {note} {The frozen matrix approach can be also implemented by using
  the fluctuation formula inside a cube, with frozen boundaries. We have
  confirmed that the fluctuation formula and the explicit deformation produce
  identical results for the shear modulus. For the bulk modulus, large pressure
  fluctuations result in a lower value when using the fluctuation formula with
  frozen boundaries.}\BibitemShut {Stop}%
\bibitem [{\citenamefont {Plimpton}(1995)}]{plimpton1995fast}%
  \BibitemOpen
  \bibfield  {author} {\bibinfo {author} {\bibfnamefont {S.}~\bibnamefont
  {Plimpton}},\ }\href@noop {} {\bibfield  {journal} {\bibinfo  {journal} {J
  Comput Phys}\ }\textbf {\bibinfo {volume} {117}},\ \bibinfo {pages} {1}
  (\bibinfo {year} {1995})}\BibitemShut {NoStop}%
\bibitem [{\citenamefont {Xu}\ \emph {et~al.}(2012)\citenamefont {Xu},
  \citenamefont {Wittmer}, \citenamefont {Poli\ifmmode~\acute{n}\else
  \'{n}\fi{}ska},\ and\ \citenamefont {Baschnagel}}]{Xu2012impulsive}%
  \BibitemOpen
  \bibfield  {author} {\bibinfo {author} {\bibfnamefont {H.}~\bibnamefont
  {Xu}}, \bibinfo {author} {\bibfnamefont {J.~P.}\ \bibnamefont {Wittmer}},
  \bibinfo {author} {\bibfnamefont {P.}~\bibnamefont
  {Poli\ifmmode~\acute{n}\else \'{n}\fi{}ska}}, \ and\ \bibinfo {author}
  {\bibfnamefont {J.}~\bibnamefont {Baschnagel}},\ }\href@noop {} {\bibfield
  {journal} {\bibinfo  {journal} {Phys. Rev. E}\ }\textbf {\bibinfo {volume}
  {86}},\ \bibinfo {pages} {046705} (\bibinfo {year} {2012})}\BibitemShut
  {NoStop}%
\end{thebibliography}%

\end{document}